\documentclass[Journal]{IEEEtran}
\IEEEoverridecommandlockouts

\usepackage{graphicx}
\usepackage{algorithm2e}
\usepackage{listings}
\usepackage{hyperref}
\usepackage{tikz}
\usepackage{xspace}
\usepackage{booktabs}
\usepackage{multicol}
\usepackage{multirow}
\usepackage{array}
\usepackage{soul}
\usepackage{subcaption}
\usepackage{enumitem}
\usepackage{amsmath,amssymb,amsfonts, amsthm} 
\usepackage{xcolor}
\usepackage{pifont}             
\usepackage{balance}
\usepackage[most]{tcolorbox}

\hypersetup{ colorlinks=true, linkcolor=black, urlcolor=black, filecolor=black, linkbordercolor=white, citecolor=black}

\SetKwComment{Comment}{/* }{ */}
\LinesNumbered
\SetKwInput{KwInput}{Input}
\SetKwInput{KwOutput}{Output}
\SetKwInput{KwParam}{Parameter}
\SetKwComment{Comment}{/*\hspace{2ex}}{*/}

\SetCommentSty{mycommfont}
\RestyleAlgo{algoruled}

\newcommand{\fram}{PoSyn\xspace}

\newtcolorbox{mybox}{
enhanced,
boxrule=0pt,frame hidden,
borderline west={4pt}{0pt}{green!75!black},
colback=green!10!white,
sharp corners
}

\begin{document}
%
\title{PoSyn: Secure Power Side-Channel Aware Synthesis}

\author{
    \IEEEauthorblockN{
    Amisha Srivastava\IEEEauthorrefmark{1},
    Samit S. Miftah\IEEEauthorrefmark{1},
    Hyunmin Kim\IEEEauthorrefmark{2},
    Debjit Pal\IEEEauthorrefmark{3},
    Kanad Basu\IEEEauthorrefmark{1}
    } \vspace{2mm}\\
    \IEEEauthorblockA{\IEEEauthorrefmark{1}ECE Department, University of Texas at Dallas, USA
    }\\
    \IEEEauthorblockA{\IEEEauthorrefmark{2}Technology Innovation Institute, UAE
    }\\
    \IEEEauthorblockA{\IEEEauthorrefmark{3}ECE Department, University of Illinois Chicago, USA
    }
}

\maketitle

%
\IEEEpeerreviewmaketitle
\begin{abstract}
Power Side-Channel (PSC) attacks exploit power consumption patterns to extract sensitive information, posing risks to cryptographic operations crucial for secure systems. Traditional countermeasures, such as masking, face challenges like complex synthesis integration, high area overhead, and vulnerability to optimization removal during logic synthesis. To address these issues, we introduce \fram, a novel logic synthesis framework designed to enhance cryptographic hardware's resistance against PSC attacks. Our approach focuses on the optimal bipartite mapping of vulnerable RTL components to standard cells from the technology library to minimize PSC leakage. By employing a cost function that integrates key characteristics from the RTL design and the standard cell library, we strategically modify the mapping criteria during the conversion of RTL designs into standard cell netlists without altering the design functionality. \textcolor{black}{Furthermore, \fram is theoretically shown to minimize mutual information leakage, further reinforcing its security against PSC vulnerabilities.} \fram is evaluated on a variety of cryptographic hardware, including AES, RSA, PRESENT, and post-quantum cryptography algorithms like Saber and CRYSTALS-Kyber across 65nm, 45nm, and 15nm nodes. \textcolor{black}{Our experimental results demonstrate a significant reduction of success rates for Differential Power Analysis (DPA) and Correlation Power Analysis (CPA) attacks, as low as 3\% and 6\%, respectively. Furthermore, TVLA analysis confirms that the synthesized netlists exhibit negligible leakage.} Moreover, compared to traditional countermeasures such as masking and shuffling, \fram achieves notably lowers the success rates, achieving a reduction by up to 72\%, while simultaneously enhancing area efficiency by as much as 3.79$\times$. These results highlight the effectiveness of \fram in securing cryptographic hardware with minimal impact on area and performance. 
\end{abstract}

\section{Introduction}
\label{Introduction}

Power Side-Channel (PSC) attacks represent a formidable threat to cryptographic systems, leveraging observable physical phenomena such as dynamic power consumption to covertly extract sensitive information from encryption hardware. The advent of side-channel attacks has highlighted significant security vulnerabilities, where adversaries exploit measurable physical effects during cryptographic operations. Central to this concern is the ability of these attackers to use such phenomena to breach the security of crucial computational processes \cite{randolph2020power}. These attacks have a wide range of effects, impacting a diverse array of cryptographic algorithms, including but not limited to symmetric and asymmetric key algorithms, hash functions, and digital signature schemes \cite{smart2000physical}. As these cryptographic algorithms are fundamental to securing network protocols, enhancing their resilience against PSC attacks is crucial for maintaining the trustworthiness of these systems and ensuring the confidentiality, integrity, and availability of sensitive information across various applications.

To address these challenges, a spectrum of countermeasures has been developed to safeguard sensitive data by obfuscating observable effects. Most existing approaches for mitigating PSC attacks are implemented at the post-silicon level \cite{huss2013amasive, wang2012scare, hwang2006aes, schmidt2009probing}. These include techniques such as real-time PSC attack detection systems that employ on-chip sensors for threat analysis and test vector leakage assessment \cite{becker2013test, gattu2020power}. A major limitation of conducting PSC analysis at the post-silicon level is the difficulty in retrofitting security measures onto existing devices, which often results in expensive device redesigns. Therefore, the need to combat PSC attacks at the pre-silicon stage is paramount. Intervening early empowers designers to embed security features directly into the hardware design, ensuring robust defenses immediately. This early-stage intervention is important, as it offers a more amenable and cost-effective environment to modify hardware designs compared to rectifying security vulnerabilities in deployed systems. 


At the Register Transfer Level (RTL), masking schemes with robust theoretical foundation have garnered significant attention as effective PSC countermeasures  \cite{akkar2001implementation, blomer2004provably, golic2003multiplicative, oswald2005side}. Masking involves splitting sensitive information into random shares and processing these separately to dilute any exploitable patterns in power consumption \cite{messerges2000securing}. It aims to diffuse sensitive information across multiple components, making it more challenging for attackers to reconstruct the original data from observable side-channel emissions. While the theoretical principles behind masking are sound, the practical implementation during the synthesis phase, which translates RTL designs into a netlist for manufacturing, introduces several challenges. The complexity added by integrating masking techniques often results in a significant increase in overhead where the final design can be multiple times larger than the original \cite{belaid2015masking}. Moreover, practical challenges such as glitches or transition-based leakages compromise the critical assumption of independent leakage between shares, rendering the secure implementation of masking both challenging and time-consuming \cite{balasch2015cost}. Furthermore, in the optimization phase of logic synthesis, which is focused on refining the design for efficiency and cost-effectiveness, tools may remove what they perceive as redundant logic. This can 
inadvertently weaken or completely strip away the protective measures introduced by masking, leaving the final netlist susceptible to attacks~\cite{tempelmeier2016maskver}. 

\begin{table*}[t!]
    \centering
    \renewcommand{\arraystretch}{1.3}
    \caption{\fram vs. Existing Approaches}
    \label{tab:comparison}
    \begin{tabular}{|l|c|c|c|}
        \hline
        \textbf{Approach} & \textbf{PSC Leakage Reduction} & \textbf{Area Overhead} & \textbf{Frequency Impact} \\
        \hline
        \textbf{Masking} & \checkmark Significant Reduction & \textcolor{red}{\textbf{High Overhead}} & \textcolor{red}{\textbf{Potential Degradation}} \\
        \hline
        \textbf{Conventional Synthesis} & \textcolor{red}{\textbf{Not Considered}} & \checkmark Optimized for Area & \checkmark Optimized for Frequency \\
        \hline
        \textbf{PoSyn} & \checkmark Integrated Minimization & \checkmark Optimized for Area & \checkmark Meets Timing Constraints \\
        \hline
    \end{tabular}
\end{table*}

Efforts to refine these masking schemes for enhanced efficiency and security are frequently marred by practical setbacks, including design inaccuracies and flaws \cite{moos2019glitch, knichel2021automated}. Such vulnerabilities lead to the requirement for a synthesis strategy that can reinforce defenses of the generated netlist against PSC attacks without the need to introduce extra operations in the RTL design. The traditional focus on performance and cost-efficiency during logic synthesis needs to be balanced with security, ensuring that protective measures are not compromised. Therefore, an effective logic synthesis strategy is required to obscure power consumption patterns, reinforcing the security of the generated netlist against PSC attacks without adding extra operations to the RTL design.

\textbf{In this paper, we propose, \fram, a side-channel-aware synthesis approach that strategically specifies the mapping criteria during the conversion of RTL designs into gate-level netlists.} \fram obscures power consumption patterns by generating a netlist that inherently resists power side-channel attacks while preserving the functional integrity of the original cryptographic design. \textcolor{black}{To further illustrate the advantages of our proposed approach, Table \ref{tab:comparison} presents a comparative analysis of masking, conventional synthesis, and \fram. This comparison underscores the necessity of a synthesis framework that effectively balances security, area constraints, and performance.}

The major contributions of our paper are as follows:
\begin{itemize}
\item We, for the first time, introduce \fram, a novel logic synthesis framework that enhances cryptographic hardware resilience against PSC attacks by optimizing standard cell selection for vulnerable RTL components.
  \item We theoretically prove that \fram is secure, minimizing mutual information leakage and thereby, significantly enhancing its resilience against PSC vulnerabilities.
   \item We applied \fram to various cryptographic hardware, including AES, RSA, PRESENT, and post-quantum cryptography algorithms Saber and Kyber, and evaluated it on 65nm, 45nm, and 15nm libraries. \fram reduced DPA and CPA success rates to as low as 3\% and 6\%, achieving up to 72\% improvement over conventional synthesis.
   \item Test Vector Leakage Assessment results confirm that PoSyn-generated netlists exhibit no detectable leakage across the cryptographic benchmarks and technology libraries, validating the security guarantees of \fram.

\item Compared to existing masking and shuffling schemes, \fram reduces PSC attack success rates by up to 72\%, while also achieving a 3.70$\times$ reduction in area overhead, offering a significant improvement in both security and resource efficiency.
    
\end{itemize}


\section{Related Works}
Power Side-Channel (PSC) attacks are of three types: (1) Simple Power Analysis (SPA), which observes power patterns \cite{mangard2003simple}, (2) Differential Power Analysis (DPA), which uses statistical analysis over multiple operations \cite{kocher1999differential, kocher2011introduction} and (3) Correlation Power Analysis (CPA), which correlates power consumption with intermediate values \cite{brier2004correlation}.
Both post-silicon and pre-silicon countermeasures have been explored to mitigate PSC attacks. Post-silicon techniques, such as test vector leakage assessment and structural information extraction \cite{becker2013test, gattu2020power, huss2013amasive, schmidt2009probing}, detect vulnerabilities after fabrication but face limitations due to the need for costly device respins, as discussed in Section \ref{Introduction}. 
In contrast, pre-silicon methods address PSC vulnerabilities during the design phase. Balancing power consumption aims to maintain a constant profile regardless of operations but introduces additional logic requirements \cite{fang2015balance}. Shuffling disrupts power correlations by randomizing the order of operations, though this increases the complexity and results in high area overhead \cite{veyrat2012shuffling}. Masking involves adding random values to the data before and after processing, requiring extra hardware to manage the masked operations \cite{messerges2000securing, damgaard2010perfectly}. Dual-rail precharge logic represents each bit with two wires, significantly increasing design complexity, area, and power consumption due to the doubled circuitry \cite{morrison2014synthesis}.


\section{Threat Model}
\label{sec:threat}
Our threat model follows standard PSC attack assumptions \cite{randolph2020power}. The adversary can interact with the cryptographic hardware to encrypt arbitrary plaintexts and collect corresponding power traces. However, the adversary cannot alter the netlist's internal architecture or access additional side-channel data (\textit{e.g}., timing or electromagnetic emissions). The attack is external, relying solely on observable power traces. The goal is to analyze multiple traces across different states of the algorithm in order to determine the secret cryptographic key. By leveraging Differential Power Analysis (DPA) and Correlation Power Analysis (CPA) \cite{kocher1999differential, brier2004correlation}, the adversary applies statistical techniques to correlate power fluctuations with internal computations, gradually extracting sensitive cryptographic information from repeated observations.
\section{PoSyn}
\label{sec:methodology}

\begin{figure}
    \centering
    \includegraphics[width=0.45\textwidth]{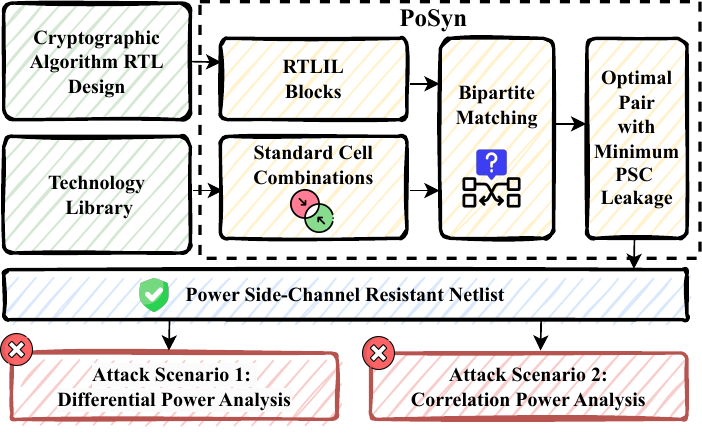}
    \caption{Overview of the proposed side-channel aware synthesis framework (PoSyn) for synthesizing a netlist using only the RTL design and technology library as inputs.}
     \label{fig:overview}
\end{figure}

The proposed approach, \fram, is designed to preserve the functionality of the RTL design while significantly obfuscating the associated power consumption patterns that could be exploited for PSC attacks. Figure \ref{fig:overview} shows an overview of \fram. It accepts the RTL design and the technology library as inputs and generates a netlist immune to PSC attacks. The approach employs two steps: (1) PSC-aware synthesis for vulnerable components, and (2) standard optimization synthesis for non-vulnerable components. Utilizing these targeted methods, \fram enhances the security of the synthesized hardware without compromising design performance.


\subsection{Vulnerable Component Identification in RTL}
\label{sec:rtl}


\textcolor{black}{\fram initiates the process by identifying operations pivotal for encryption in the RTL design of the cryptographic algorithm \cite{he2019rtl, 10508974}. Our approach is partially user-defined, allowing designers to incorporate domain knowledge and specify criteria for marking components as vulnerable. This involves recognizing complex operations that are often repeated, like those found in \textit{Sbox} lookup tables in AES. It also requires identifying sensitive variables, such as encryption keys, which are at a high risk of causing information leakage. We refine our analysis to discern the ``leaky'' modules and specific locations within these modules that are engaged in these critical operations and variables, thereby marking them as areas particularly vulnerable to PSC attacks. This strategy enables us to identify and subsequently fortify the design's most critical areas against potential leakage. The following components from the RTL design are identified for the proposed side-channel aware synthesis method.}
\begin{itemize}

    \item \underline{Sensitive variables}:  The primary focus is on protecting sensitive variables, such as encryption keys, round keys, and critical intermediate states (\textit{e.g.}, \textit{SBox} outputs in substitution-permutation ciphers). The exposure of these sensitive variables poses a dire risk, potentially compromising the security of encrypted data. %
\item \underline{Encryption-specific operations}: Core encryption operations, which are susceptible to PSC attacks due to their repetitive nature and predictable power signatures, must be protected. Safeguarding these operations helps prevent the exploitation of their power profiles.
 \item \underline{Leaky modules}: RTL modules prone to power leakage need to be identified using existing detection methods \cite{he2019rtl, 10508974}. These modules, which often handle sensitive variables or critical operations, contribute to power leakage due to bit-flipping and variable data patterns. We mitigate the risk of PSC attacks exploiting these power variations by securing these modules, such as the \textit{SubBytes} module in AES.
    \item \underline{Computationally intensive operations}: Even if not directly related to encryption, certain operations can still have a significant impact on the power profile of the hardware. Variations in power consumption during these operations can unintentionally reveal details about underlying processes, leading to indirect data leakage. By safeguarding these operations, power consumption patterns are obscured, reducing the risk of attackers exploiting these variations.

\item \underline{High-fanout components}: Components with high fanout play a crucial role in shaping the overall power distribution of the circuit. Their strategic inclusion in our protection strategy comes from the understanding that a uniform power distribution complicates the execution of PSC attacks. Addressing these components helps in masking the overall power consumption patterns, thereby obfuscating any sensitive data from potential attackers.
\end{itemize}

 
The next step in \fram is preventing these identified components from being targeted by PSC attacks.

\subsection{Conversion to RTLIL Representation}

Transitioning from RTL to RTLIL (Register Transfer Level Intermediate Language) is a crucial step in logic synthesis, facilitated by tools like Yosys \cite{wolf2016yosys}. RTLIL captures the design’s functionality and structural layout, detailing logic operations, data flows, and interconnections. This transformation breaks down RTL designs into fundamental elements such as cells, registers, and logic gates, enabling precise and direct mapping to physical components in the technology library cells while preserving design specifications. For instance, a simple half adder defined in RTL with \textit{XOR} and \textit{AND} operations for the sum and carry (Listing \ref{listing:rtl_alu}) is represented in RTLIL as \textit{XOR} and \textit{AND} cells within a module (Listing \ref{listing:rtlil_alu}). 

\definecolor{codegreen}{rgb}{0,0.6,0}
\definecolor{codegray}{rgb}{0.5,0.5,0.5}
\definecolor{codepurple}{rgb}{0.58,0,0.82}
\definecolor{backcolour}{rgb}{0.98,0.98,0.95}

\lstdefinestyle{mystyle}{
    framexleftmargin=10pt,
    frameround=tttt,
    frame=single,
    commentstyle=\color{codegreen},
    keywordstyle=\color{blue},
    numberstyle=\tiny\color{codegray},
    stringstyle=\color{codepurple},
    basicstyle=\ttfamily\footnotesize,
    breakatwhitespace=false,         
    breaklines=true,                 
    captionpos=b,                    
    keepspaces=true,                 
    numbers=left,                    
    numbersep=5pt,                  
    showspaces=false,                
    showstringspaces=false,
    showtabs=false,                  
    tabsize=1,
    xleftmargin= 15pt,
    xrightmargin= 5pt
}

\lstset{style=mystyle}
\begin{lstlisting}[language=Verilog, caption={RTL Code for half adder.},label={listing:rtl_alu}] 
module half_adder (
    input wire a, b,      
    output wire sum, carry);
assign sum = a ^ b; assign carry = a & b;
endmodule
\end{lstlisting}

\lstset{style=mystyle}
\begin{lstlisting}[language=Verilog, caption={RTLIL Code for half adder.},label={listing:rtlil_alu}] 
module \half_adder
  wire width 1 input 0 \a, input 1 \b
  wire width 1 output 2 \sum, output 3 \carry
  cell $xor $xor_cell
    parameter \A_SIGNED 0
    parameter \B_SIGNED 0
    parameter \Y_WIDTH 1
    connect \A \a
    connect \B \b
    connect \Y \sum
  end
  cell $and $and_cell
    parameter \A_SIGNED 0
    parameter \B_SIGNED 0
    parameter \Y_WIDTH 1
    connect \A \a
    connect \B \b
    connect \Y \carry
  end
end
\end{lstlisting}




\subsection{Cell Selection from Technology Library}

\begin{algorithm}[h]
\caption{Standard Cell Mapping (SCM)}
\label{algo:scm}
\KwInput{$RTLIL\_Block$}
\KwOutput{$Optimal\_Cell\_Config$}
$SC \leftarrow \text{Get\_Std\_Cell\_Lib}()$\; 
$Func \leftarrow \text{Extract\_Func}(RTLIL\_Block)$\; 
$Direct\_Matches \leftarrow \text{Find\_Direct}(Func, SC)$\; 
$Indirect\_Comb \leftarrow []$\; 
$Comp\_Funcs \leftarrow \text{Decompose}(Func)$\; 
\For{$func \in Comp\_Funcs$}{
    $Indirect\_Matches \leftarrow \text{Explore\_Indirect}(func, SC)$\; 
    $Indirect\_Comb.append(Indirect\_Matches)$\;
}
$All\_Matches \leftarrow Direct\_Matches + Indirect\_Comb$\; 
$Optimal\_Config \leftarrow \text{Simulated\_Annealing}(All\_Matches)$\; 
\Return $Optimal\_Config$
\end{algorithm}



Once the RTL design is translated into RTLIL, the next step maps vulnerable components to standard cells from the technology library.  In this section, we focus on generating all the possible cell combinations for RTLIL blocks of the vulnerable RTL components. Once the combinations are selected, we prioritize the combination, that minimizes PSC leakage, (further elaborated in Section \ref{sec:psc}). Algorithm \ref{algo:scm} offers an overview of the approach for the selection of the combinations. Initially, the algorithm retrieves the available standard cells and extracts the specific functionality of the RTLIL block (lines 1-2). It then conducts a path exploration, searching for direct matches of the RTLIL cell functionality within the standard cell library (line 3). It also simultaneously decomposes the RTLIL cell functionality into simpler components to explore indirect mappings (lines 5-8). These potential solutions, encompassing both direct matches and indirect combinations, are then merged into a unified set (line 10). Finally, the algorithm employs simulated annealing to find an optimal solution from this set based on the defined criteria (line 11).


\subsubsection{Generating Combinations}
We employ a structured methodology to accurately translate the functionality of each RTLIL block by selecting standard cells from the given technology library. Our approach is twofold:
\begin{itemize}
    \item \textbf{Direct Mapping}: Direct Mapping involves an immediate search for standard cells within the technology library that exactly match the functionality specified by the RTLIL block. For example, for an RTLIL block designed to execute a NAND operation, the step involves cataloging all NAND gate cells available in the standard cell library to ensure a broad spectrum of direct functional matches.
    \item \textbf{Indirect Mapping}: Indirect mapping decomposes the RTLIL block's functionality into basic logical components, such as \textit{AND}, \textit{OR}, and \textit{XOR} gates, and explores combinations of simpler standard cells to emulate the original functionality. This hierarchical decomposition and recombination process breaks down complex operations, like arithmetic or combinational logic, into manageable elements for efficient mapping. By deconstructing these functionalities into their constituent parts, we increase the number of potential configurations. Each configuration represents a unique combination of simpler standard cells joined together to emulate the original, more complex function. This exploration spans various combinations of logical gates within the confines of the standard cell library, adhering to a well-defined threshold for combination exploration.
\end{itemize}

 

\textcolor{black}{We set a threshold for exploring the various combinations to efficiently navigate the expansive search space while maintaining a balance between functional accuracy and area overhead. This threshold constrains the exploration to combinations that involve a limited number of standard cells, ensuring that area overhead and power leakage remain within acceptable bounds. This threshold is determined through an optimization process guided by simulated annealing, as detailed in Section \ref{sec:simulated}.}

\subsubsection{Simulated Annealing to choose combinations}
\label{sec:simulated}


In the optimization process for selecting standard cell combinations to accurately replicate RTLIL block functionalities, simulated annealing plays a crucial role in navigating the complex solution space \cite{bertsimas1993simulated, kirkpatrick1983optimization}. This method efficiently searches through a large set of potential combinations, identifying the ones that best fulfill the synthesis goals while managing complexity. By using a probabilistic technique, simulated annealing systematically refines the selection of standard cell combinations, ensuring that they match the desired functionality and adhere to constraints such as the number of cells used, which helps to control the circuit size and power consumption.

Simulated Annealing begins with an initial set of combinations, including both direct and indirect mappings of RTLIL blocks to corresponding standard cells. Through iterative exploration, simulated annealing generates ``neighboring" solutions by slightly altering the current combination: adding, removing, or substituting cells. Each generated combination is evaluated against a threshold that limits the maximum allowable number of standard cells per combination. This threshold is either predefined based on area and power constraints or adaptively adjusted within the simulated annealing process to balance optimization efficiency with design feasibility. The criterion concerning the number of standard cells serves as a filter during this evaluation, disqualifying combinations that exceed the set limit.
As the simulated annealing process progresses, the algorithm fine-tunes its exploration, increasingly favoring combinations that closely align with the optimization goals. It reduces the likelihood of accepting suboptimal combinations over time, thereby focusing the search on solutions that satisfy the criteria, including the limitation on the number of cells.
The algorithm concludes with a selection of optimal combinations that replicate the intended RTLIL functionalities and also adhere to the design constraints imposed by area overhead and static power leakage. Following the selection of these combinations, we proceed to choose the optimal combination based on criteria that minimize the PSC leakage during synthesis. This has been elaborated subsequently in Section \ref{sec:psc}.

\subsection{Optimal Bipartite Matching for PSC-Aware Mapping}
\label{sec:psc}
\noindent
Finally, with potentially multiple valid combinations for each RTLIL block, the process culminates in the selection of a single optimal mapping for each block. This selection is achieved by utilizing bipartite matching, which considers the entire set of RTLIL blocks and their corresponding valid combinations \cite{karp1990optimal}. The goal is to find the best overall mapping that minimizes the circuit's PSC leakage according to a set of predefined criteria while maintaining functional correctness.
 The following steps highlight the construction process of the Bipartite graph:
 
\noindent
 1. \textbf{Set Formation}: The bipartite graph is constructed with two distinct sets of vertices: one representing the RTLIL cells (Set 1) and the other representing the combinations of standard cells (Set 2). 

\noindent
2.\textbf{ Edge Creation}: Edges are drawn between vertices from Set 1 to Set 2, where each edge represents a potential mapping from an RTLIL cell to a standard cell combination. 
Functional equivalence between the RTLIL cells and standard cell combinations is validated through graph isomorphism techniques to ensure feasibility of these connections.

\noindent
3. \textbf{Cost function}: The cost function plays a pivotal role in determining the optimal mapping of RTLIL cells to standard cell combinations. This function is crafted from a mixture of criteria derived from the RTL design and standard cell library data, focusing on minimizing PSC leakage, described as follows:

\color{black}
\subsubsection{Factors Derived from Standard Cell Library}

From the standard cell library, we derive information crucial for mitigating PSC leakage:

\begin{itemize}
    \item \textbf{Driving Strength (DS)}: The driving strength of a cell influences the speed of signal transitions and peak current during switching. Higher driving strength can increase power signatures, making PSC attacks more effective, but it is necessary for high fanout cases to maintain integrity.

    \item \textbf{Capacitance (C)}: The capacitance of a cell affects the power required for switching. Higher capacitance results in increased dynamic power consumption, thus raising the risk of power side-channel leakage. Reducing capacitance helps minimize power variations for cells involved in complex or frequent operations.
\end{itemize}

\subsubsection{Factors Influencing PSC Leakage from the RTL}

Power consumption leakage in digital circuits is influenced by multiple factors, including the nature of computations and the structural characteristics of the circuit components. These factors include:

\begin{itemize}
    \item \textbf{Sensitive Variables (SV)}:  For RTLIL blocks containing sensitive variables, such as encryption keys, mapping these blocks to standard cells with lower driving strengths reduces PSC leakage risks by minimizing power variations. Lower driving strengths lead to slower transitions and reduced peak currents, thus diminishing detectable power signatures exploitable in attacks. However, it is essential to ensure that these driving strengths do not fall below a threshold that might compromise the circuit’s frequency or induce voltage drops causing functional issues. For example, in AES encryption, round keys are mapped to cells with lower driving strengths, limiting power fluctuations while maintaining performance.
    

\lstset{style=mystyle}
\begin{lstlisting}[language=Verilog, caption={RTLIL Code with complex operation.},label={listing:rtlil2}] 
module SubBytes
  wire width 8 input 0 data_in
  wire width 8 output 1 data_out
  memory 256 sbox
  init sbox {
    \63 \7c \77 \7b \f2 \6b \6f \c5 \30 \01 \67
    // Remaining S-box entries
  }
    assign data_out = sbox[data_in]
  end
end
\end{lstlisting}

    \item \textbf{Intensive Operations (IO)}: For RTILIL cells with complex and computationally intensive operations, the cost function favors mappings to cells with low capacitance and sufficient driving strength to balance power consumption and signal integrity. Lower capacitance minimizes dynamic power consumption, while higher driving strength ensures signal integrity during intricate operations, preventing delays that could increase static power consumption. For example, in AES the \textit{SubBytes} step uses an \textit{SBox} to perform non-linear byte substitution on the state matrix. The RTLIL code snippet in Listing \ref{listing:rtlil2} demonstrates the implementation of this \textit{SubBytes} step, which leads to a peak in the power profile of the algorithm.
    
    

    \item  \textbf{High Fanout Components (F)}: 
    For high-fanout components not directly involved in encryption, mapping to standard cells with higher driving strengths supports large loads efficiently, while introducing intentional variability in the power profile to obscure sensitive data processing. By selectively adjusting the power profile in these regions, PSC analysis can be misled, thereby reducing the risks of information leakage. Listing \ref{listing:rtlil3} provides an example of high-fanout operations in the AES \textit{MixColumns}.

\end{itemize}

\lstset{style=mystyle}
\begin{lstlisting}[language=Verilog, caption={RTLIL Code showcasing high fanout components.},label={listing:rtlil3}] 
module MixColumns
  wire width 8 input data_in0, data_in1, data_in2, data_in3
  wire width 8 output data_out0, data_out1, data_out2, data_out3
  assign data_out0 = data_in0 ^ data_in1; 
  assign data_out1 = data_in1 ^ data_in2;
  assign data_out2 = data_in2 ^ data_in3; 
  assign data_out3 = data_in3 ^ data_in0; 
end
\end{lstlisting}

\begin{figure}[t!]
    \centering
    \includegraphics[width=\linewidth]{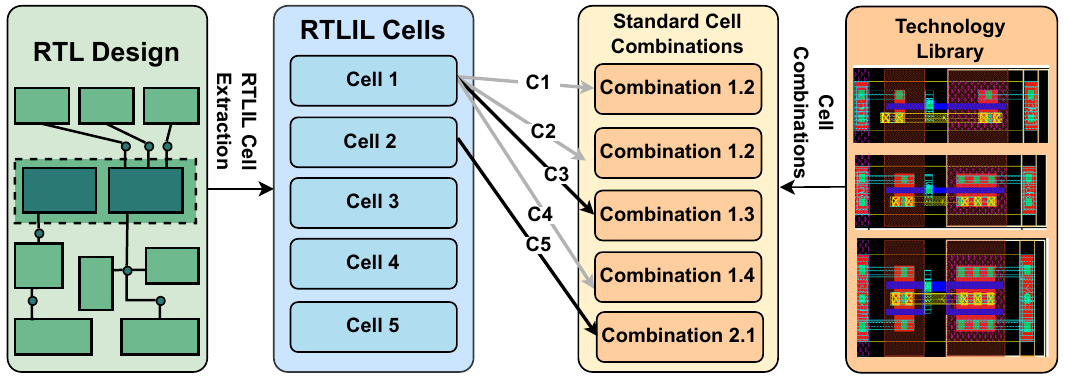}
    \caption{Optimal Bipartite Matching: The grey lines from Set 1 (RTLIL cells) to Set 2 (standard cell combinations) indicate all possible mappings and the black lines indicate mappings with the lowest value of the cost function C. 
    The optimal matching with the lowest C 
    is selected for each vertex from Set 1.} 
     \label{fig:bipartite}
     \vspace{-5mm}
\end{figure}


\subsubsection{\textbf{Derivation of the Cost Function}}
To minimize PSC leakage, the cost function must account for relationships between these factors and their impact on power consumption. 

The weighting factors \( \alpha, \beta, \gamma \) are introduced to appropriately scale the impact of each term based on its significance. These factors are determined through empirical methods such as grid search or optimization based on design-specific power and performance requirements. This flexibility allows the cost function to adapt to various design constraints, ensuring effective minimization of PSC leakage. 
\\
\noindent
\textbf{Term \#1: Sensitivity to Switching Activity and Driving Strength:} For an RTL component \( A \) with sensitive variables, the risk of leakage increases with higher switching activity. To mitigate this, we prioritize minimizing power variations by controlling the switching activity using driving strength \( DS \). A higher switching activity can lead to more pronounced power fluctuations while \( DS \) influences the peak current during transitions. Therefore, we include a term inversely proportional to \( DS \) to prioritize higher driving strength cells, which can handle frequent transitions more effectively:
\begin{equation}
\text{Term \#1:} \quad \frac{\alpha \cdot SV}{DS}
\end{equation}
Here:
\begin{itemize}
    \item \( SV \) is a binary indicator (1 if the block contains sensitive variables, 0 otherwise).
    \item The term \( \frac{1}{DS} \) balances higher switching activity with the need for robust handling of frequent transitions.
\end{itemize}

\noindent
\textbf{Term \#2: Impact of Number of Operations and Capacitance:}

For each RTL component  \( A \), the dynamic power consumption in its mapped standard cell  \( S \) is directly influenced by the number of intensive operations performed and the capacitance \( C \). As the number of operations increases, the switching activity \( SA \) also tends to increase because more signal transitions are likely to occur. Thus, the number of intensive operations can be considered a proxy for switching activity in the cost function.

We capture this relationship by including a term that combines the number of operations \( IO \) and the capacitance \( C \). Higher capacitance results in increased dynamic power consumption, especially when the number of operations (and thus switching activity) is high:
\begin{equation}
\text{Term \#2:} \quad \beta \cdot IO \cdot C
\end{equation}
Here:
\begin{itemize}
    \item \( IO \) represents the number of intensive operations performed by the block, which correlates with higher switching activity.
    \item \( C \) is the capacitance of the cell, affecting the power consumed during switching.
\end{itemize}
This term ensures that blocks performing a higher number of operations, especially those with higher capacitance, are accounted for in the cost function to reduce power consumption variations.

\noindent
\textbf{Term \#3: Influence of Fanout and Driving Strength:}
For an RTL component \(A\) with high fanout, its mapped standard cell \(S\) requires adequate driving strength to drive large loads effectively. While higher driving strength can increase power consumption, it is necessary to ensure signal integrity when handling large fanouts.
\begin{equation}
\text{Term \#3:} \quad \gamma \cdot F \cdot DS
\end{equation}
Here:
\begin{itemize}
    \item \( F \) represents the fanout or the number of loads driven by the signal.
    \item \( DS \) is the driving strength of the cell.
\end{itemize}
The term reflects the need to balance adequate signal strength with efficient power management for high fanout scenarios.

\noindent
\textbf{Complete Cost Function:}  
Combining the derived terms, the overall cost function becomes:
\begin{equation}
C (A, S) = \sum_{i} \left( \alpha \cdot \frac{SV}{DS_i} + \beta \cdot IO \cdot C_i + \gamma \cdot F \cdot DS_i \right)
\label{eq:cost}
\end{equation}

Here, \( C (A, S)\) denotes the cost of mapping the RTL component A to the standard cell S and \( i \) indexes the standard cells (in Set 2). 

Minimizing this cost function leads to selecting standard cell combinations for every RTLIL block that achieves an optimal balance between ensuring functional accuracy and minimizing PSC leakage from the design, as shown in Figure \ref{fig:bipartite}. The cost function equation is designed to be universally applicable across all pairs of vertices in Set 1 (RTLIL blocks) and Set 2 (standard cell combinations) for the mapping process. \textbf{With the bipartite graph constructed and the cost function defined, an optimal bipartite matching algorithm, the Hungarian algorithm, is employed to find the set of mappings that minimizes the total cost \cite{fukuda1992finding}.} This process ensures that each RTLIL cell is mapped to the most suitable standard cell combination, taking into account the sensitive nature of the variables involved, the computational demands of the design, and the overall goal of achieving a power side-channel resistant mapping.

Finally, when all the RTLIL cells have been mapped to a combination of standard cells, we obtain the PSC-resistant netlist components. The rest of the RTLIL cells (which were not identified as prone to PSC attacks) are synthesized with the usual criteria in order to optimize area, power, and performance. In this manner, the generated netlist is not only resistant to PSC attacks but also fulfills the optimization needed for the placement and routing step.

\section{Post-Synthesis Verification for Functional Correctness}
\textcolor{black}{Once the PSC-resistant netlist is generated, we check its functional correctness using a Logic Equivalence Checker (LEC) tool, between the RTL design and the post-synthesis netlist generated using \fram. The LEC tool, Synopsys Formlaity, is used in standard industry hardware design flow, post logic-synthesis, to ensure the functional correctness of the netlist \cite{formality2010equivalence}. The tool evaluates whether the logical functions of the post-synthesis netlist are identical to those of the original RTL, thus ensuring that the synthesis process has not altered the intended functionality. By confirming both structural and functional equivalence, we ensure that the generated netlist is the true functional equivalent of the RTL design. }


\section{\textcolor{black}{Theoretical Analysis of Posyn's Security Guarantee}}

As mentioned in Section \ref{sec:psc}, \fram employs the Hungarian algorithm to derive an optimal mapping \( M \) that minimizes the cost function \( C \), representing power consumption leakage associated with cryptographic operations. By leveraging this cost-minimization approach, \fram effectively reduces the quantity of side-channel information that can be inferred from power traces, thus improving the system's resilience to PSC attacks. 

\smallskip


\noindent
\textbf{Mutual Information and Leakage Reduction:} Inheriting the concept of mutual information gain from information theory, we evaluate how much knowledge about a cryptographic system can be extracted from observed side-channel leakage. In this scenario, the mutual information gain \( I(K, L) \) measures the dependency between the cryptographic key \( K \) and the leakage \( L \). It is defined as:

\begin{equation}
I(K, L) = H(K) - H(K \mid L),
\end{equation}
where:
\begin{itemize}
    \item \( H(K) \) is the entropy of the key, quantifying 
    its initial uncertainty.
    \item \( H(K \mid L) \) is the conditional entropy of the key given the observed leakage, quantifying 
    the remaining uncertainty about the key after observing \( L \).
\end{itemize}



The goal of \fram is to reduce the correlation between power consumption and the cryptographic key. This is achieved by minimizing the cost function 
\(C(M)\), defined as: \begin{equation} C(M) = \sum_{(A, S)} C(A, S) \end{equation} where 
\(C(A,S)\) denotes the cost of mapping the RTL component \(A\) to the standard cell \(S\), as specified in Equation \ref{eq:cost}.

Since power side-channel leakage \(L\) arises from variations in power consumption across different operations, it depends on the underlying physical implementation of the design. Specifically, the structural mapping of RTL components to standard cells affects switching activity, driving strength, and capacitance, all of which influence power consumption. Therefore, we model leakage as a function of the total mapping cost: \begin{equation} L = f(C(M)) \end{equation} where \(f(C(M))\) captures the dependency between the design’s structural mapping and its corresponding power leakage characteristics.


After minimizing the cost function \( C(M) \) directly impacts the mutual information \( I(K, L) \) by controlling design aspects linked to power leakage. 
Based on derivation of Equation \ref{eq:cost}:
\begin{itemize}
    \item Term \#1 minimizes fluctuations by balancing switching activity with driving strength, reducing power variations caused by sensitive operations, and decreasing the correlation between power traces and key-dependent activity.
    \item Term \#2 addresses high-capacitance cells and intensive operations, suppressing dynamic power signatures that could otherwise reveal patterns linked to K.
    \item Term \#3 optimizes driving strength in high-fanout cases, stabilizing power and preventing leaks that could indicate specific operational patterns.
\end{itemize}
 Together, these terms increase the conditional entropy \( H(K|L) \), effectively reducing \( I(K, L) \) and making the power traces more independent of the cryptographic key. \newline


\noindent
\textbf{Impact on Entropy and Mutual Information:}
The Hungarian algorithm guarantees that the mapping \( M \) is optimal for minimizing \( C(M) \), directly impacting \( I(K, L) \) by reducing power-based leakage sources. As a result of the targeted reductions in \( C(M) \), PoSyn increases \( H(K \mid L) \) such that:
\begin{equation}
H(K | L) \to H(K) \Rightarrow I(K, L) \to 0
\end{equation}


Consequently, as \( I(K, L) \to 0 \), the side-channel leakage conveys negligible information regarding the cryptographic key, significantly enhancing resilience against power-based side-channel attacks. 

\textcolor{black}{By reasoning 
that \( I(K, L) \to 0 \), we establish that the leakage \( L \) exhibits near-independence from the secret cryptographic key \( K \). This reduction in mutual information serves as a theoretical guarantee for the enhanced security of PoSyn, affirming the Hungarian algorithm’s efficacy in constructing a mapping resistant to PSC attacks.}



\section{Results}
\label{sec:results}

\subsection{Experimental Setup}

We use cryptographic RTL design benchmarks to perform our side-channel aware synthesis and validate the efficacy of our approach. We evaluate \fram across three technology libraries: 65nm, 45nm, and 15nm \cite{UMC65nm, Nangate45, martins2015open}. We also utilize the LEC tool, Synopsys Formality, to perform post-synthesis verification on the synthesized netlists \cite{formality2010equivalence}. First, simulation is performed to collect power measurements for both the PoSyn-generated netlists and the conventional netlists synthesized by Yosys (since it is a widely used open-source synthesis tool for conventional synthesis). Please note that any other synthesis tool can also be used for the same. For each cryptographic hardware implementation, we simulated the design with a wide range of input values while collecting corresponding power measurements. These values were recorded under identical operational conditions for both the side-channel resistant and conventional netlists, ensuring a fair basis for comparison. The tools utilized for this process were Synopsys VCS for the Value Change Dump (VCD) file generation and PrimeTime for the power measurements. Two sets of experiments are presented in this section: (1) DPA attack, and (2) CPA attack. In each experiment, 4000 power traces were captured and analyzed.


\subsubsection{Benchmarks}
We conduct an evaluation of \fram on four distinct implementations of the AES algorithm, namely, \textit{AES\_Comp}, \textit{AES\_TBL}, \textit{AES\_PPRM1}, and \textit{AES\_PPRM3} \cite{aes}. We also evaluate our method on the encryption algorithms RSA and PRESENT, with \textit{RSA1024\_RAM} and \textit{PRESENT} benchmarks, respectively \cite{aes, present}. Furthermore, we also evaluate two lattice-based Post Quantum Cryptography (PQC) algorithms: Saber and CRYSTALS-Kyber \cite{imran2021design, yaman2021hardware}. 

\subsubsection{Metrics for Comparison}

To evaluate the effectiveness of \fram in mitigating PSC leakage, we consider the following two key metrics:


\begin{itemize}

\item \textbf{Success Rate:} The effectiveness of DPA and CPA attacks depends on their ability to accurately extract secret keys through power measurement analysis. In DPA, this involves generating key hypotheses, testing possible values, and refining them based on observed power consumption patterns. In CPA, each key guess is evaluated using a predictive model that estimates power consumption, typically leveraging the Hamming weight of data due to its strong correlation with power usage.

The success rate of these attacks serves as a crucial metric, offering a quantitative measure of their efficacy against cryptographic algorithms. Defined as the proportion of successful key extractions to the total attempts made, this metric is represented by Equation \ref{eq:res}.

\begin{equation} \label{eq:res}
\text{ Success Rate} = \frac{\text{No. of Successful Key Recoveries}}{\text{Total No. of Attack Attempts}}
    \end{equation}


Here, ``Total No. of Attack Attempts" refers to the total number of attempts or iterations an attacker undertakes to successfully extract the secret key or significant portions of it from the cryptographic algorithm. 
\textcolor{black}{In each iteration, the attack is conducted using a new set of power traces, and if the extracted candidate key matches the true key, that iteration is considered a success. This metric provides a statistical measure of how reliably an attack can recover the key across multiple independent attempts. A lower success rate indicates that the attacker needs many attempts to extract the key, while a higher success rate suggests that the key can be reliably recovered with fewer iterations.}

\item \textbf{Test Vector Leakage Assessment:} 
Test Vector Leakage Assessment (TVLA) is a statistical method used to evaluate the presence of power side-channel leakage in cryptographic implementations \cite{ding2018towards}. This metric provides a quantitative measure of how distinguishable power traces are when subjected to different input conditions, offering insights into the effectiveness of countermeasures against PSC attacks.

TVLA is conducted by comparing power traces obtained using fixed versus random input vectors. Welch’s t-test is applied to each time sample to calculate t-values, which indicate the level of statistical disparity between the two sets. If the cryptographic implementation exhibits no leakage, the distribution of t-values should be approximately normal under the null hypothesis \cite{ding2016simpler}. 

Consequently, a standard threshold of ±4.5 corresponds to an extremely low probability (on the order of 0.00001\%) of observing such a value purely by chance \cite{schneider2016leakage}. This means that in the absence of leakage, the probability of a t-value exceeding ±4.5 is about one in ten million. Any t-value surpassing this threshold is considered a strong indicator of exploitable leakage \cite{ferrufino2023fobos}. A lower proportion of t-values exceeding ±4.5 suggests greater resilience against PSC attacks, whereas higher occurrences indicate statistical evidence of leakage.

\end{itemize}

Next, we proceed to calculate the success rates for the netlists synthesized by \fram (side-channel resistant) and the netlists generated by Yosys (conventional synthesis). A lower success rate indicates that the attack required numerous attempts to successfully extract the key. \textcolor{black}{Following this, we present the TVLA results for all benchmarks across the three technology libraries. These results provide a statistical validation of power leakage, complementing the success rate analysis by quantifying how distinguishable the power traces are under different input conditions.}

\subsection{Success Rate Results}

Table \ref{tab:my-table} summarizes the success rates of DPA and CPA attacks for the various cryptographic benchmarks across three technology libraries: UMC 65nm, Nangate 45nm, and Nangate 15nm. For each benchmark, the table presents the DPA and CPA success rates for the side-channel resistant netlists generated by \fram, along with the percentage change in the synthesized netlist compared to the conventional netlist. It can be observed that across all three libraries, DPA has a lower success rate compared to CPA. In our evaluation of \fram for cryptographic cores, the success rate is influenced by the limited number of power measurements available for analysis. This constraint reflects real-world scenarios where attackers may face technical or access-related limitations. In DPA, having a limited number of power measurements makes it challenging to conduct the statistical analysis required to accurately distinguish the key-dependent variations from noise.


Moreover, Figures \ref{fig:DPA_main} and \ref{fig:CPA_main} illustrate the DPA and CPA success rates for \fram-generated side-channel resistant netlists and conventional netlists across all benchmarks for the three synthesis libraries. Each graph corresponds to a specific synthesis library, with the x-axis representing the benchmarks, while the y-axis shows the success rates of the attacks for both types of netlists.
\begin{table*}[t!]
\centering
\caption{Benchmark evaluation results on success rates of DPA and CPA for the encryption algorithms along with \% change in netlist in terms of standard cells for three technology libraries.}
\label{tab:my-table}
\resizebox{\textwidth}{!}{%
\begin{tabular}{|c|ccc|lll|lll|}
\hline
\multicolumn{1}{|l|}{} &
  \multicolumn{3}{c|}{65 nm} &
  \multicolumn{3}{c|}{45 nm} &
  \multicolumn{3}{c|}{15 nm} \\ \hline
Benchmark &
  \multicolumn{2}{c|}{Success Rate} &
  \multirow{2}{*}{\begin{tabular}[c]{@{}c@{}}Change \\ in Netlist\end{tabular}} &
  \multicolumn{2}{c|}{Success Rate} &
  \multicolumn{1}{c|}{\multirow{2}{*}{\begin{tabular}[c]{@{}c@{}}Change \\ in Netlist\end{tabular}}} &
  \multicolumn{2}{c|}{Success Rate} &
  \multicolumn{1}{c|}{\multirow{2}{*}{\begin{tabular}[c]{@{}c@{}}Change \\ in Netlist\end{tabular}}} \\ \cline{1-3} \cline{5-6} \cline{8-9}
 &
  \multicolumn{1}{c|}{\begin{tabular}[c]{@{}c@{}}Differential \\ Power \\ Analysis\end{tabular}} &
  \multicolumn{1}{c|}{\begin{tabular}[c]{@{}c@{}}Correlation \\ Power\\ Analysis\end{tabular}} &
   &
  \multicolumn{1}{c|}{\begin{tabular}[c]{@{}c@{}}Differential \\ Power \\ Analysis\end{tabular}} &
  \multicolumn{1}{c|}{\begin{tabular}[c]{@{}c@{}}Correlation \\ Power\\ Analysis\end{tabular}} &
  \multicolumn{1}{c|}{} &
  \multicolumn{1}{c|}{\begin{tabular}[c]{@{}c@{}}Differential \\ Power \\ Analysis\end{tabular}} &
  \multicolumn{1}{c|}{\begin{tabular}[c]{@{}c@{}}Correlation \\ Power\\ Analysis\end{tabular}} &
  \multicolumn{1}{c|}{} \\ \hline
AES\_Comp &
  \multicolumn{1}{c|}{0.05} &
  \multicolumn{1}{c|}{0.08} &
  19\% &
  \multicolumn{1}{l|}{0.04} &
  \multicolumn{1}{l|}{0.08} &
  21\% &
  \multicolumn{1}{l|}{0.03} &
  \multicolumn{1}{l|}{0.06} &
  25\% \\ \hline
AES\_TBL &
  \multicolumn{1}{c|}{0.10} &
  \multicolumn{1}{c|}{0.13} &
  15\% &
  \multicolumn{1}{l|}{0.09} &
  \multicolumn{1}{l|}{0.16} &
  17\% &
  \multicolumn{1}{l|}{0.08} &
  \multicolumn{1}{l|}{0.12} &
  20\% \\ \hline
AES\_PPRM1 &
  \multicolumn{1}{c|}{0.15} &
  \multicolumn{1}{c|}{0.20} &
  21\% &
  \multicolumn{1}{l|}{0.12} &
  \multicolumn{1}{l|}{0.18} &
  24\% &
  \multicolumn{1}{l|}{0.12} &
  \multicolumn{1}{l|}{0.17} &
  30\% \\ \hline
AES\_PPRM3 &
  \multicolumn{1}{c|}{0.12} &
  \multicolumn{1}{c|}{0.18} &
  20\% &
  \multicolumn{1}{l|}{0.10} &
  \multicolumn{1}{l|}{0.18} &
  22\% &
  \multicolumn{1}{l|}{0.10} &
  \multicolumn{1}{l|}{0.15} &
  28\% \\ \hline
RSA1024\_RAM &
  \multicolumn{1}{c|}{0.16} &
  \multicolumn{1}{c|}{0.21} &
  25\% &
  \multicolumn{1}{l|}{0.07} &
  \multicolumn{1}{l|}{0.13} &
  28\% &
  \multicolumn{1}{l|}{0.12} &
  \multicolumn{1}{l|}{0.18} &
  35\% \\ \hline
PRESENT &
  \multicolumn{1}{c|}{0.10} &
  \multicolumn{1}{c|}{0.15} &
  10\% &
  \multicolumn{1}{l|}{0.08} &
  \multicolumn{1}{l|}{0.10} &
  12\% &
  \multicolumn{1}{l|}{0.08} &
  \multicolumn{1}{l|}{0.12} &
  18\% \\ \hline
SABER &
  \multicolumn{1}{c|}{0.09} &
  \multicolumn{1}{c|}{0.12} &
  27\% &
  \multicolumn{1}{l|}{0.13} &
  \multicolumn{1}{l|}{0.22} &
  29\% &
  \multicolumn{1}{l|}{0.08} &
  \multicolumn{1}{l|}{0.10} &
  33\% \\ \hline
KYBER &
  \multicolumn{1}{c|}{0.14} &
  \multicolumn{1}{c|}{0.18} &
  22\% &
  \multicolumn{1}{l|}{0.11} &
  \multicolumn{1}{l|}{0.19} &
  20\% &
  \multicolumn{1}{l|}{0.10} &
  \multicolumn{1}{l|}{0.15} &
  30\% \\ \hline
\end{tabular}%
}
\end{table*}


\subsubsection{\textbf{Results: UMC 65nm Library}}
The UMC 65nm library is designed for advanced ASIC development utilizing UMC's 65 nm Low-K Standard Performance SHVT process and offers 216 cell types with multiple drive strengths to support high-density IC designs \cite{UMC65nm}.

\begin{figure*}[t!]
     \centering
     \begin{subfigure}[b]{0.33\textwidth}
         \centering
         \includegraphics[width=\textwidth]{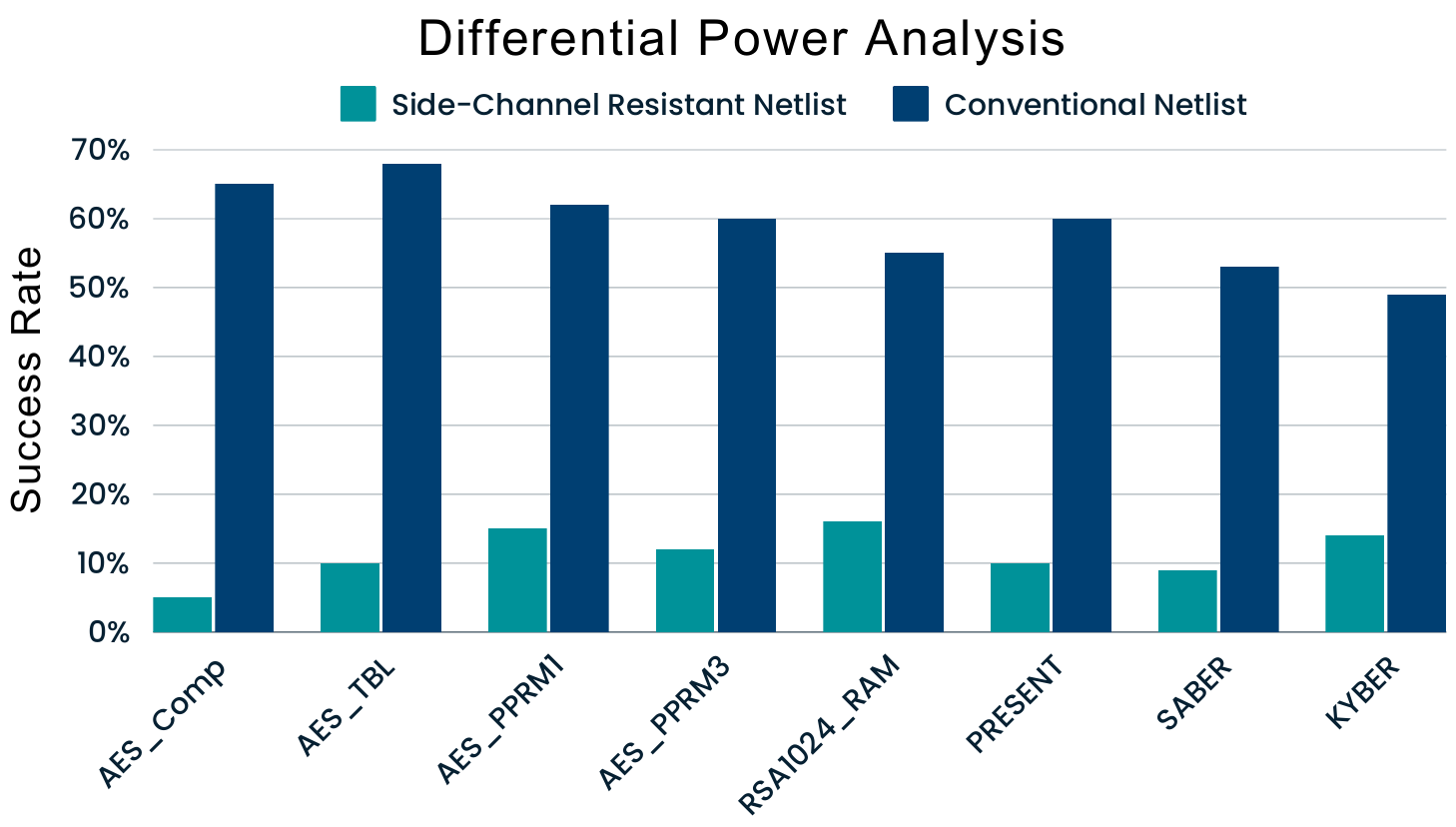}
         \caption{65nm Library}
         \label{fig:dpa}
     \end{subfigure}
     \hfill
     \begin{subfigure}[b]{0.328\textwidth}
         \centering
         \includegraphics[width=\textwidth]{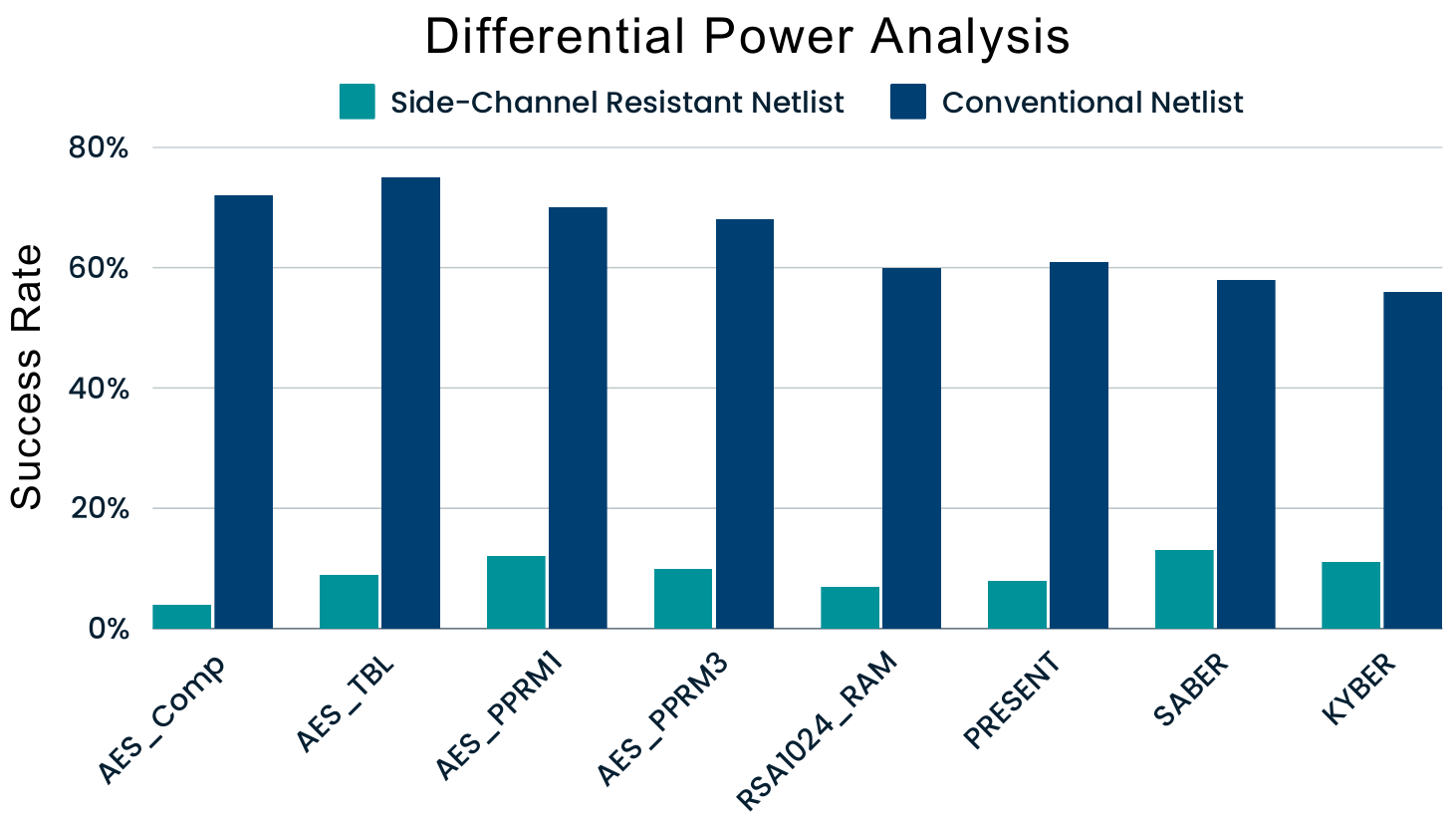}
         \caption{45nm Library}
         \label{fig:dpa2}
     \end{subfigure}
     \hfill
     \begin{subfigure}[b]{0.33\textwidth}
         \centering
         \includegraphics[width=\textwidth]{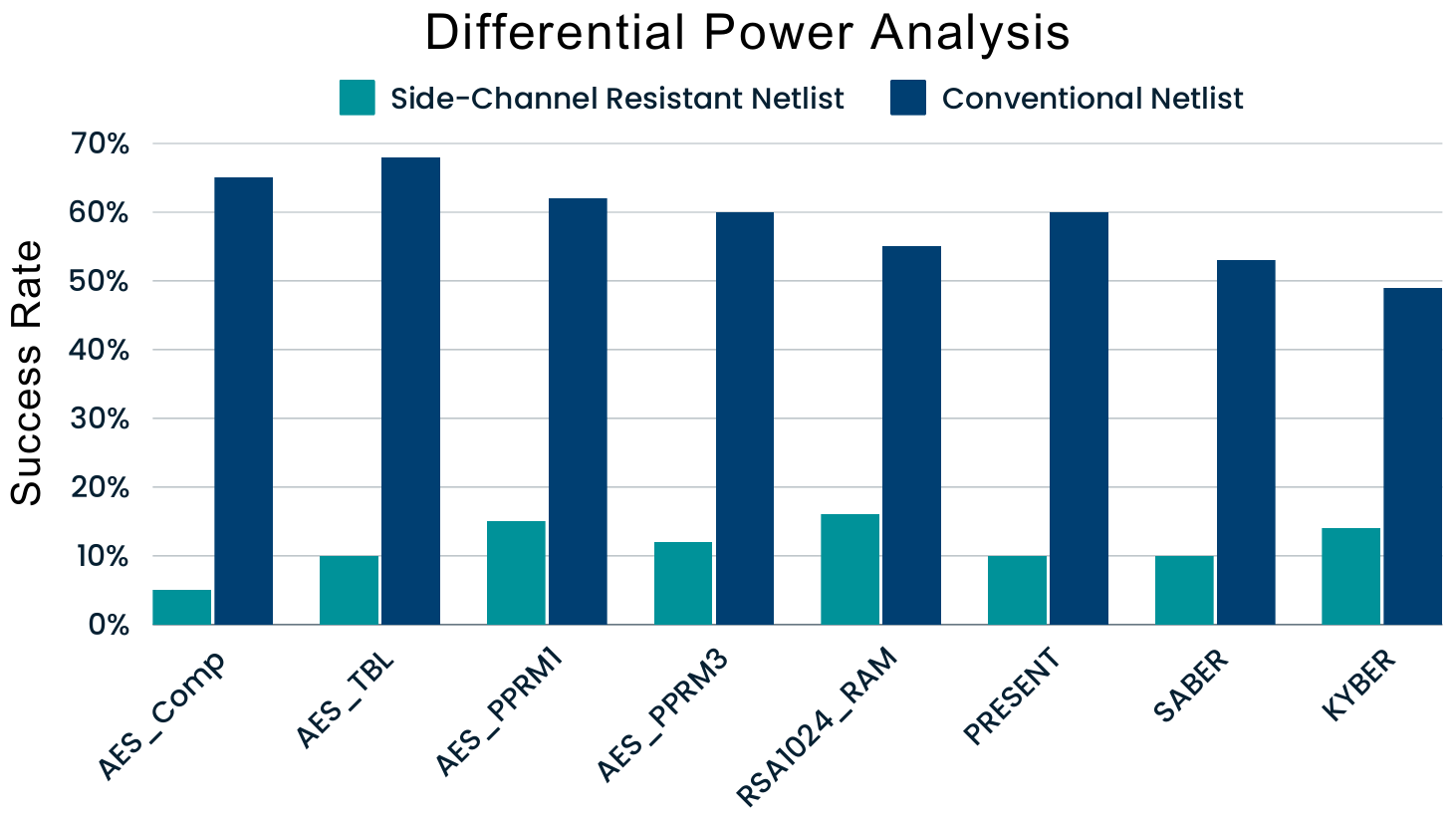}
         \caption{15nm Library}
         \label{fig:dpa1}
     \end{subfigure}\
        \caption{DPA results for Side-Channel Resistant vs Conventional Netlist: Graphs showcase the success rates for the attack across all benchmarks for the three synthesis libraries.}
\label{fig:DPA_main}
\end{figure*}

\begin{figure*}[t!]
     \centering
     \begin{subfigure}[b]{0.33\textwidth}
         \centering
         \includegraphics[width=\textwidth]{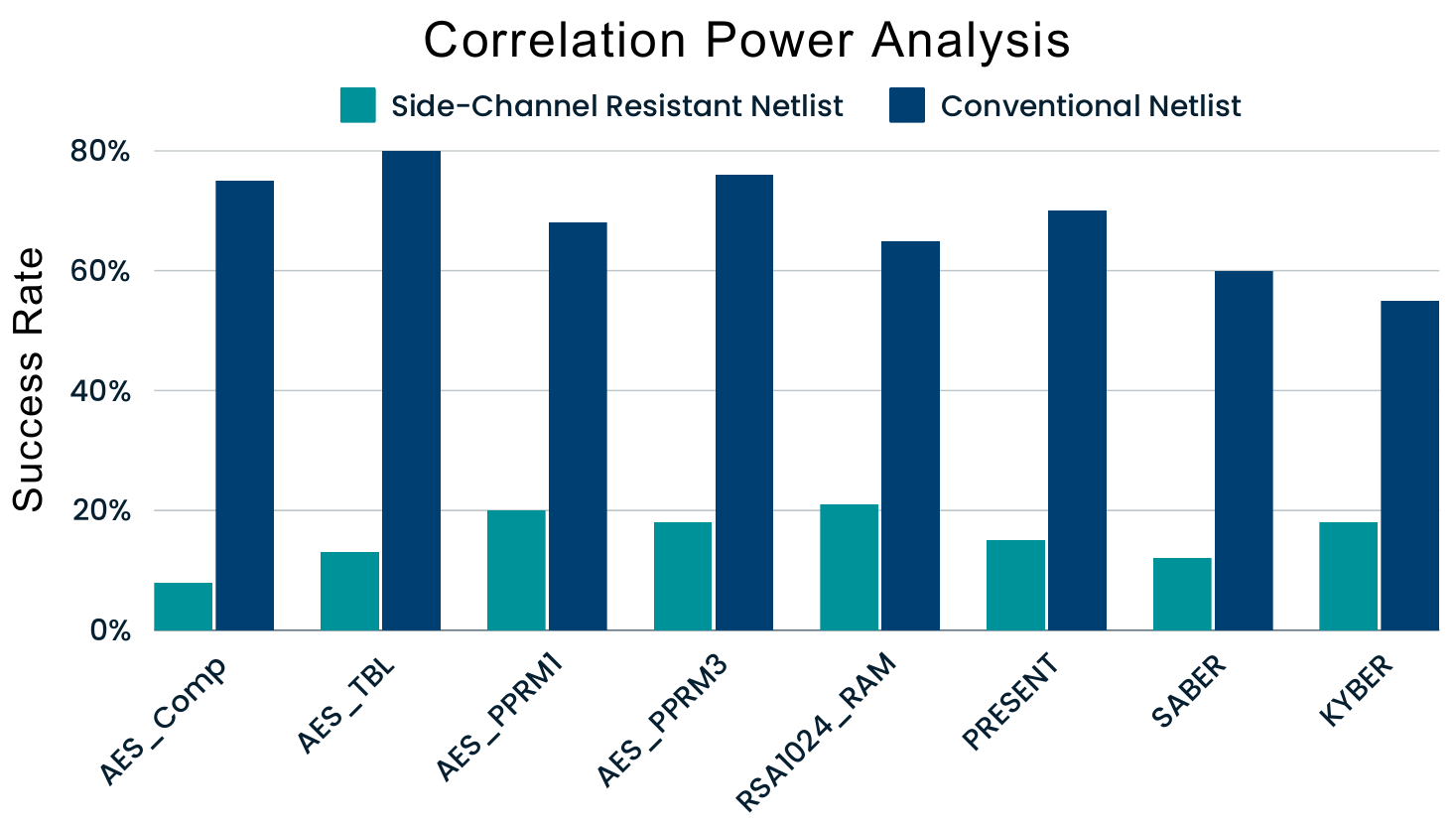}
         \caption{65nm Library}
         \label{fig:cpa}
     \end{subfigure}
     \hfill
     \begin{subfigure}[b]{0.328\textwidth}
         \centering
         \includegraphics[width=\textwidth]{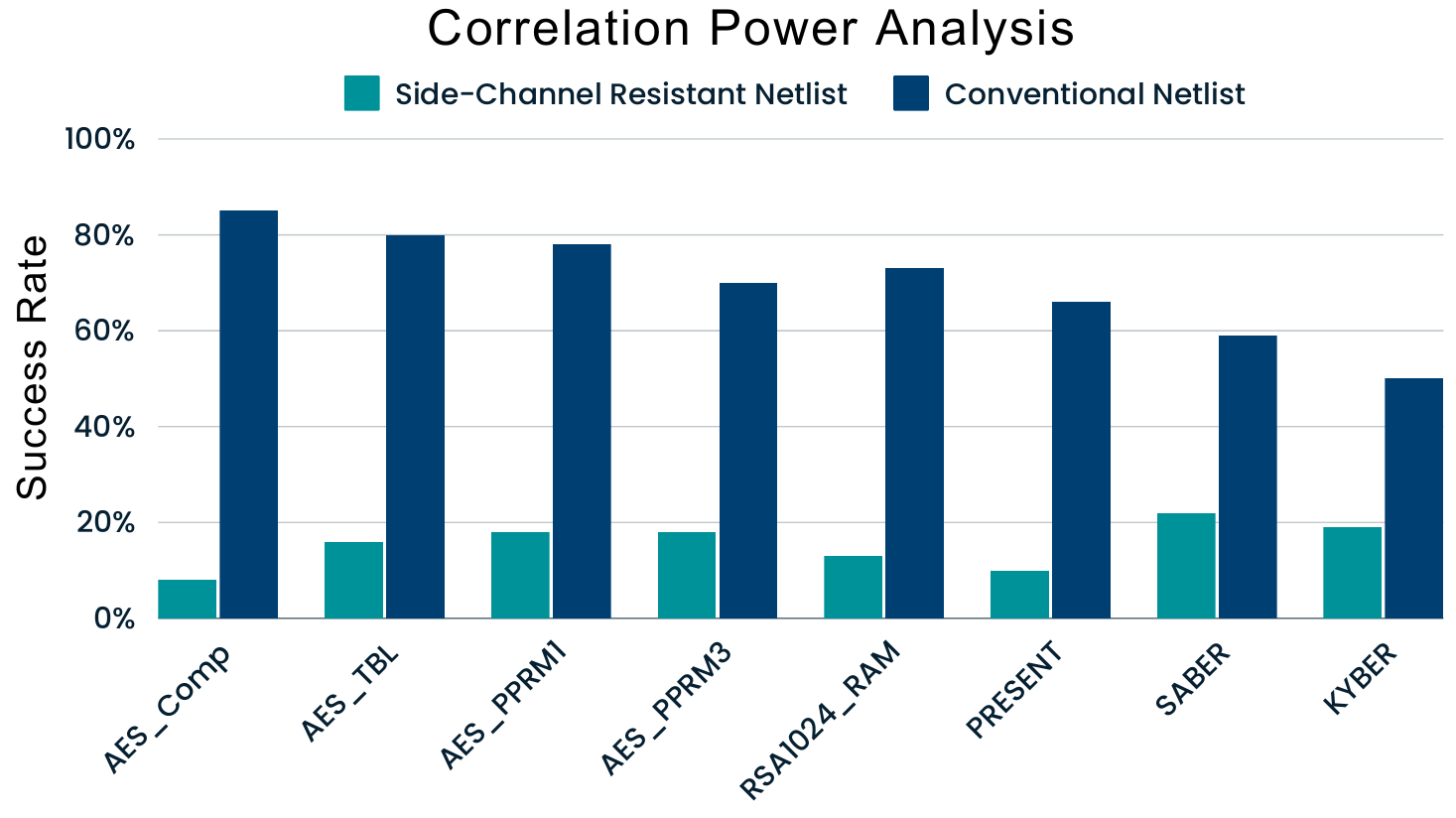}
         \caption{45nm Library}
         \label{fig:cpa2}
     \end{subfigure}
     \hfill
     \begin{subfigure}[b]{0.33\textwidth}
         \centering
         \includegraphics[width=\textwidth]{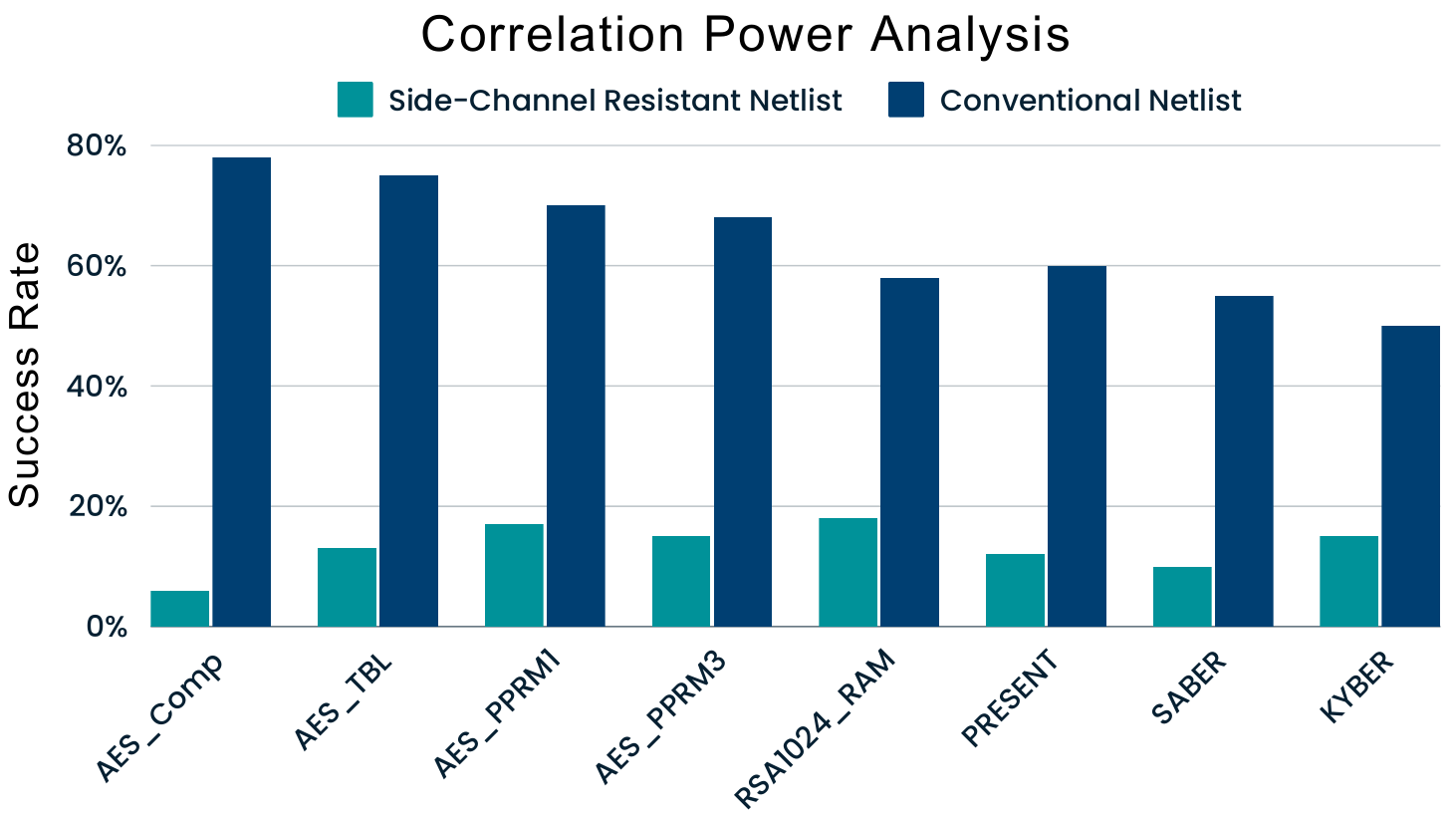}
         \caption{15nm Library}
         \label{fig:cpa1}
     \end{subfigure}
     \caption{CPA results for Side-Channel Resistant vs Conventional Netlist: Graphs showcase the success rates for the attack across all benchmarks for the three synthesis libraries.}
\label{fig:CPA_main}
\end{figure*}

The results shown for the 65nm library in Table \ref{tab:my-table} demonstrate that the side-channel resistant netlists achieve significantly lower success rates for both DPA and CPA attacks, with modest changes in the netlist structure, highlighting the effectiveness of \fram in enhancing security while maintaining design efficiency. 
Figures \ref{fig:dpa} and \ref{fig:cpa} showcase a comparison of success rates between the side-channel resistant netlist synthesized by our framework and the conventional netlist synthesized by Yosys. It can be observed that the success rate values are significantly lower for the side-channel resistant netlists generated by \fram across all benchmarks. For the \textbf{\textit{AES\_Comp} benchmark, the success rate for a DPA attack on the side-channel resistant netlist is remarkably low at 5\%}, compared to a much higher 65\% success rate for the conventional netlist, achieving a maximum reduction in DPA success rate of up to 60\% with only a 19\% change in the netlist, which reflects the percentage of cells that differ from the conventional netlist, indicating the proportion of cell types that were either added or replaced in the transformed netlist.

Moreover, as seen in Figure \ref{fig:cpa}, \textbf{when shifting to CPA, the success rate for the \textit{AES\_Comp} benchmark slightly increases to 8\% for the side-channel resistant netlist}, yet it remains significantly lower than the 75\% success rate observed for the conventional netlist.
This increase in the success rate from DPA to CPA reflects the greater precision of CPA in exploiting correlations within power consumption data, yet it also highlights the effectiveness of the method in keeping the success rate low. The trend of increased success rates for CPA attacks in comparison to DPA attacks persists across all benchmarks. Nonetheless, the side-channel resistant netlists consistently exhibit substantially lower success rates than their conventional counterparts for both types of attacks. Even for benchmarks involving asymmetric cryptography algorithms such as \textit{RSA\_1024} and PQC schemes like \textit{SABER} and \textit{KYBER}, the side-channel resistant netlists showcase a lower success rate. For instance, \textit{RSA\_1024} benchmark shows a success rate of 16\% for DPA and 21\% for CPA, which are significantly lower than the 55\% and 65\% rates for the conventional netlist, respectively. Moreover, with a significantly high number of power measurements, the success rate of the conventional netlist without any countermeasures can go up to 100\%. Lastly, it can be observed that \fram incurs at most 27\% change in the netlist cells.

\subsubsection{\textbf{Results: Nangate 45nm Open Cell Library}}
The NanGate 45nm Open Cell Library, designed for educational and research use, offers a versatile set of 134 types of standard cells each with different driving strengths, that are crafted to align with the demands of the 45nm process \cite{Nangate45}.

Utilizing this library, \fram continues to demonstrate significant improvements in mitigating side-channel attacks across various cryptographic benchmarks, as seen in Table \ref{tab:my-table}. The success rates for DPA and CPA on the side-channel resistant netlists are consistently lower than those for the conventional netlists, as seen in Figures \ref{fig:dpa2} and \ref{fig:cpa2}. \textbf{It can be observed that the \textit{AES\_Comp} benchmark shows a DPA success rate of only 4\% for the side-channel resistant netlist}, compared to a much higher rate of 70\% for the conventional netlist, resulting in a maximum reduction in DPA success rate of up to 66\% with PoSyn.
Similarly, \textbf{the success rate for CPA on the \textit{AES\_Comp} benchmark for the side-channel resistant netlist is 8\%}, which is still substantially lower than the 78\% observed for the conventional netlist, achieving a maximum reduction in CPA success rate of up to 70\%.
This trend is consistent across all benchmarks. Additionally, the side-channel resistant netlists consistently exhibit a significant percentage change in netlist structure, with changes ranging from 12\% to 29\%, depending on the benchmark. Notably, benchmarks involving asymmetric cryptography algorithms such as \textit{RSA\_1024} and post-quantum cryptography schemes like \textit{SABER} and \textit{KYBER} also demonstrate substantially lower success rates on the side-channel resistant netlists. 



\subsubsection{\textbf{Results: NanGate 15nm Open Cell Library}}
The NanGate 15nm Open Cell Library, developed with Silvaco and based on NCSU’s FreePDK15, offers 76 digital cell types with varying drive strengths, supporting the demands of modern 15nm process technologies \cite{martins2015open}. By utilizing 15nm technology, \fram has enhanced the resistance of cryptographic netlists to PSC attacks. The reduced transistor size minimizes power consumption and signal leakage, making DPA and CPA attacks more challenging. Table \ref{tab:my-table} reflects the efficacy of our side-channel resistant designs: the success rates for DPA and CPA are considerably lower than those of conventional netlists. 
\textbf{For example, in the \textit{AES\_Comp} benchmark, the side-channel resistant netlist by \fram achieves a maximum reduction in success rates of 65\% for DPA and 66\% for CPA compared to the conventional netlist.} This notable reduction in success rates persists across various cryptographic algorithms, demonstrating our method's robustness in mitigating risks of key extraction. Other algorithms like \textit{RSA\_1024} and post-quantum cryptography algorithms such as \textit{SABER} and \textit{KYBER} also showcase significant reductions in success rates, underscoring the benefits of our security enhancements at the 15nm scale. This has been illustrated in Figures \ref{fig:dpa1} and \ref{fig:cpa1} showing Posyn's effectiveness in significantly lowering the success rates of the attacks.

\begin{tcolorbox}[colback=blue!5!white,colframe=blue!75!black] \textbf{Summary}: Older nodes are inherently more vulnerable due to higher power consumption, making them more susceptible to side-channel attacks. \fram mitigates these vulnerabilities effectively, achieving consistently low success rates with modest netlist changes across all benchmarks. \end{tcolorbox}

\subsection{Test Vector Leakage Assessment}

\begin{figure}[t!]
    \centering
    \includegraphics[width=\linewidth]{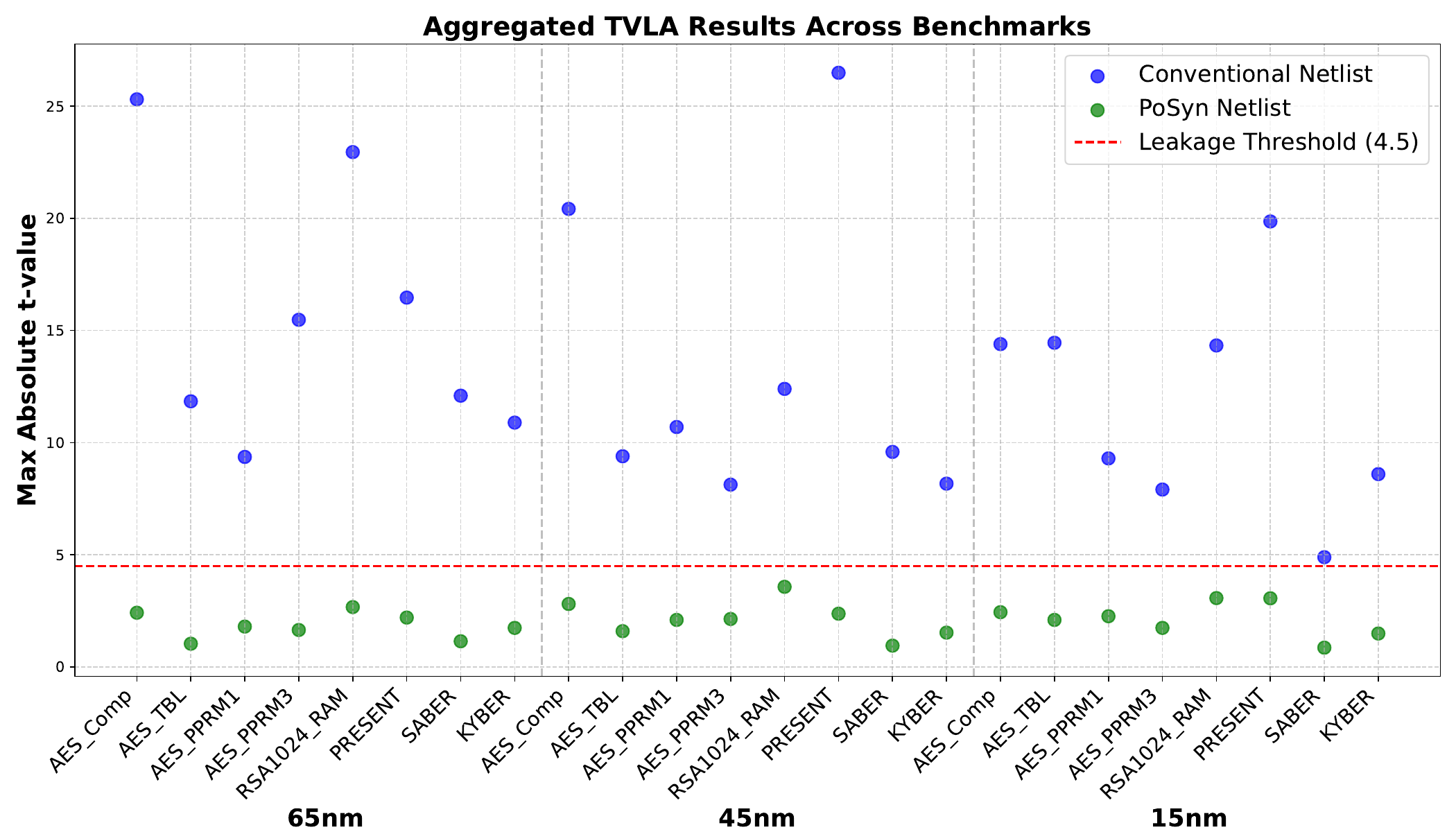}
    \caption{ Maximum absolute t-values from TVLA for cryptographic benchmarks across 65nm, 45nm, and 15nm nodes.} 
     \label{fig:tvla}
     
\end{figure}

Figure \ref{fig:tvla} illustrates the results of the Test Vector Leakage Assessment (TVLA) for multiple cryptographic benchmarks synthesized using conventional and PoSyn methodologies across the three technology libraries. The x-axis represents various benchmarks, grouped by technology node, while the y-axis shows the maximum absolute t-value observed for each benchmark-library combination. 

It can be observed that the conventional netlists exhibit widespread leakage, with maximum absolute t-values consistently exceeding the standard threshold across all benchmarks and technology libraries, confirming their vulnerability to DPA and CPA attacks. In contrast, the \fram-generated netlists exibit t-values below the standard threshold for all tested benchmarks and libraries, demonstrating significantly reduced PSC leakage. 

These results indicate that \fram's synthesis-level modifications successfully prevent data-dependent power variations, making cryptographic hardware more resilient against PSC attacks. The effectiveness of PoSyn is consistent across multiple cryptographic algorithms including AES, RSA, PRESENT, SABER, and KYBER, as well as across different technology nodes, underscoring its robustness as a scalable and technology-independent technique.


\subsection{{Timing and Memory Overheads}}

\begin{table}[b!]
\centering
\caption{Timing and Memory Overheads Across 15nm, 45nm, and 65nm Libraries.}
\label{table:overhead}
\resizebox{\columnwidth}{!}{%
\begin{tabular}{|c|c|cc|}
\hline
\multirow{2}{*}{\textbf{Library}} & \multirow{2}{*}{\textbf{Benchmark}} & \multicolumn{2}{c|}{\textbf{Overhead}} \\ \cline{3-4} 
 &  & \multicolumn{1}{c|}{Timing Overhead} & Memory Overhead \\ \hline
\multirow{8}{*}{15nm} & AES\_Comp & \multicolumn{1}{c|}{12\%} & 10\% \\ \cline{2-4} 
 & AES\_TBL & \multicolumn{1}{c|}{7\%} & 5\% \\ \cline{2-4} 
 & AES\_PPRM1 & \multicolumn{1}{c|}{17\%} & 13\% \\ \cline{2-4} 
 & AES\_PPRM3 & \multicolumn{1}{c|}{14\%} & 7\% \\ \cline{2-4} 
 & RSA1024\_RAM & \multicolumn{1}{c|}{22\%} & 15\% \\ \cline{2-4} 
 & PRESENT & \multicolumn{1}{c|}{6\%} & 4\% \\ \cline{2-4} 
 & SABER & \multicolumn{1}{c|}{16\%} & 12\% \\ \cline{2-4} 
 & KYBER & \multicolumn{1}{c|}{13\%} & 11\% \\ \hline
\multirow{8}{*}{45nm} & AES\_Comp & \multicolumn{1}{c|}{10\%} & 8\% \\ \cline{2-4} 
 & AES\_TBL & \multicolumn{1}{c|}{6\%} & 5\% \\ \cline{2-4} 
 & AES\_PPRM1 & \multicolumn{1}{c|}{14\%} & 12\% \\ \cline{2-4} 
 & AES\_PPRM3 & \multicolumn{1}{c|}{12\%} & 8\% \\ \cline{2-4} 
 & RSA1024\_RAM & \multicolumn{1}{c|}{18\%} & 14\% \\ \cline{2-4} 
 & PRESENT & \multicolumn{1}{c|}{5\%} & 4\% \\ \cline{2-4} 
 & SABER & \multicolumn{1}{c|}{13\%} & 10\% \\ \cline{2-4} 
 & KYBER & \multicolumn{1}{c|}{10\%} & 9\% \\ \hline
\multirow{8}{*}{65nm} & AES\_Comp & \multicolumn{1}{c|}{8\%} & 7\% \\ \cline{2-4} 
 & AES\_TBL & \multicolumn{1}{c|}{5\%} & 4\% \\ \cline{2-4} 
 & AES\_PPRM1 & \multicolumn{1}{c|}{12\%} & 10\% \\ \cline{2-4} 
 & AES\_PPRM3 & \multicolumn{1}{c|}{10\%} & 7\% \\ \cline{2-4} 
 & RSA1024\_RAM & \multicolumn{1}{c|}{16\%} & 13\% \\ \cline{2-4} 
 & PRESENT & \multicolumn{1}{c|}{4\%} & 3\% \\ \cline{2-4} 
 & SABER & \multicolumn{1}{c|}{11\%} & 9\% \\ \cline{2-4} 
 & KYBER & \multicolumn{1}{c|}{9\%} & 8\% \\ \hline
\end{tabular}%
}
\end{table}

\textcolor{black}{PoSyn’s standard cell mapping prioritizes PSC resistance over traditional metrics such as power, area, and performance while ensuring that timing constraints are met. To accommodate signal propagation delays introduced by security-driven transformations, The \fram generated netlist operates at a lower clock frequency while ensuring that paths meet setup and hold timing requirements within an extended clock period. This approach maintains timing integrity while addressing performance overhead concerns. However, the reduction in clock frequency may decrease throughput, lowering the number of bits encrypted per second compared to the original design. This security-focused strategy enhances PSC resistance but introduces a trade-off, requiring users to balance security and performance.}

\textcolor{black}{Timing overhead, quantified as the percentage increase in critical path delay compared to the conventional netlist, is measured using OpenSTA. This reflects the reduction in maximum operating frequency due to PoSyn’s security-driven modifications. Additionally, while PoSyn does not explicitly optimize for memory efficiency, we assess its Memory Overhead to quantify the computational costs of storing transformed netlists. Here, Memory Overhead refers to the additional memory usage incurred by the synthesis tool when processing the PoSyn-modified design, ensuring that resource demands remain practical for larger circuits.}

\textcolor{black}{Table \ref{table:overhead} summarizes both timing and memory overheads for benchmarks across the three standard cell libraries. Notably, PoSyn's overhead remains minimal compared to existing masking or shuffling-based countermeasures, making it more efficient for protecting cryptographic cores.}


\begin{tcolorbox}[colback=blue!5!white,colframe=blue!75!black]
 \textbf{Summary}: \fram satisfies all the timing constraints without negative slack or timing violations, preserving the design's timing integrity. The memory overhead introduced by PoSyn is minimal across all evaluated libraries, ensuring efficient resource utilization while enhancing PSC resistance. 
\end{tcolorbox}

\subsection{Comparison with Existing Countermeasures}

Among widely adopted pre-silicon PSC countermeasures, masking and shuffling are the most common techniques. This section compares \fram with these methods, examining their effectiveness against PSC attacks.

\subsubsection{Masking Schemes}


Masked logic reduces data-dependent power consumption by dividing sensitive data into multiple independent shares \cite{masking}. Masking techniques often incur significant area and performance overhead due to the additional circuitry required. In contrast, \fram strategically modifies the synthesis process to enhance PSC resistance without extensive circuit modifications. To evaluate this, we implemented AES with first-order masking, using a masked \textit{Sbox} to obscure the relationship between the cryptographic key and power consumption \cite{shahmirzadi2021new}. As shown in Table \ref{tab:counter}, first-order masking reduces DPA and CPA success rates to 15\% and 18\%, respectively. \fram alone achieves a lower success rate of 3\% for DPA and 6\% for CPA. In contrast, a hybrid approach combining masking with \fram increases DPA and CPA success rates to 5\% and 8\%. These findings demonstrate PoSyn's effectiveness in generating PSC-resistant netlists compared to masking or a hybrid approach.




\textcolor{black}{Furthermore, we also evaluate the area overhead of various first-order masking schemes for AES using the 45nm library. This comparison highlights the silicon area utilization of different approaches. The area is quantified in gate equivalences (GE), normalized to the area of a 2-input NAND gate, which serves as the base unit in the given standard cell library and typically defines the technology-dependent unit area in contemporary CMOS technologies. 
Table \ref{table:compare} illustrates the comparison results and consists of three columns: the first column indicates the approach evaluated, and the second column specifies a measure of the silicon area consumed by each scheme in kilo gate equivalences (kGE). Lastly, the third column quantifies the extent of area improvement  \fram offers compared to the approach. The results are based on synthesized netlists for each scheme, revealing that \fram demonstrates the least area overhead of 4.51kGE and offers a significant improvement over other existing methods. This underscores the effectiveness of our proposed synthesis methodology, \fram, in overcoming the major limitation posed by masking. }

\begin{table}[h!]
\caption{Comparison of implementation cost of existing masking schemes for AES.}
\label{table:compare}
\begin{tabular}{|c|c|c|}
\hline
Approach & \begin{tabular}[c]{@{}l@{}}Area in kGE\end{tabular} & \begin{tabular}[c]{@{}c@{}}Improvement\\ in Area\end{tabular}\\ \hline
\fram & 4.51 &  - \\ \hline
Changing of Guards \cite{askeland2022guarding} & 13.66 &  3.02 $\times$ \\ \hline
 d + 1 Share Masking \cite{de2016masking} & 6.68 &  1.48 $\times$ \\ \hline
Threshold Implementation \cite{bilgin2015trade} & 8.12 & 1.80 $\times$ \\ \hline
4-Share AES \cite{wegener2018first}& 7.60 &   1.68 $\times$\\ \hline
2-Share AES \cite{shahmirzadi2021re} & 7.71 &  1.70 $\times$ \\  \hline
3-Share AES \cite{sugawara20193} & 17.10 & 3.79 $\times$ \\ \hline
\end{tabular}
\end{table}

\textcolor{black}{Additionally, our results indicate that \fram achieves comparable area overheads of approximately 4.79 kGE and 5.22 kGE for 65nm and 15nm technology libraries, respectively.}

\subsubsection{Shuffling Techniques}

Shuffling is another common PSC countermeasure that mitigates vulnerabilities by randomizing the order of operations, thereby reducing the correlation between power traces and sensitive data \cite{yahya2009aes, wang2013area}. To evaluate its effectiveness, we used power traces from the open-source ASCAD database, which includes traces from a shuffled AES implementation by ANSSI \cite{ascad}. The results in Table \ref{tab:counter} show that while shuffling offers some level of protection, PoSyn’s synthesis-level enhancements further reduce the success rates of DPA and CPA attacks, highlighting its added effectiveness. Due to limited access to open-source RTL implementations for shuffled AES, we were unable to measure the area overhead for the synthesis process.

\begin{table}[t!]
\centering
\caption{Comparison with Masking and Shuffling Countermeasures on AES.}
\label{tab:counter}
\resizebox{\columnwidth}{!}{%
\begin{tabular}{|c|c|c|c|}
\hline
\textbf{Success Rate} & \textbf{PoSyn} & \textbf{First Order Masking} & \textbf{Shuffling} \\ \hline
DPA & 3\% & 15\% & 9\% \\ \hline
CPA & 6\% & 18\% & 11\% \\ \hline
\end{tabular}%
}
\end{table}

\begin{tcolorbox}[colback=blue!5!white,colframe=blue!75!black]
\textbf{Summary}: \fram outperforms traditional countermeasures such as Masking and Shuffling, achieving significantly lower success rates for DPA and CPA attacks. Additionally, PoSyn incurs minimal area overhead, making it an efficient and robust solution for cryptographic hardware in high-security environments demanding strong protection against PSC attacks.
\end{tcolorbox}

\section{Conclusion}
\label{sec:conclusion}
\balance

This paper introduces \fram, a novel side-channel aware synthesis framework that enhances cryptographic hardware resistance against PSC attacks. \fram defines mapping criteria for synthesizing RTL designs to standard cell netlists, strategically integrating characteristics from both RTL designs and technology libraries. This approach preserves functional integrity while fortifying designs against PSC attacks. Additionally, \fram is theoretically proven to minimize mutual information leakage, further reinforcing its security against PSC vulnerabilities. The framework’s effectiveness has been validated across benchmarks, including AES, RSA, PRESENT, and post-quantum algorithms like Saber and CRYSTALS-Kyber. Tested on 65nm, 45nm, and 15nm nodes, \fram consistently reduces DPA and CPA success rates, achieving 3\% and 6\%, respectively, for 15nm nodes. \textcolor{black}{Furthermore, TVLA analysis confirms that the \fram-synthesized netlists consistently maintain t-values well within the ±4.5 threshold, thereby ensuring negligible side-channel leakage.} \fram mitigates PSC vulnerabilities with minimal trade-offs: timing overheads up to 22\% and memory overheads up to 15\%. Compared to masking and shuffling techniques, \fram reduces CPA success rates and improves area efficiency by up to $3.79\times$. These results highlight PoSyn’s robustness, offering an effective solution for securing cryptographic devices in high-security applications such as network protocols and systems.

\section{acknowledgment}
This research is partially supported by Technology Innovation Institute (TII), Abu Dhabi, UAE.

\bibliographystyle{IEEEtran}
\bibliography{References}

\begin{thebibliography}{10}
\providecommand{\url}[1]{#1}
\csname url@samestyle\endcsname
\providecommand{\newblock}{\relax}
\providecommand{\bibinfo}[2]{#2}
\providecommand{\BIBentrySTDinterwordspacing}{\spaceskip=0pt\relax}
\providecommand{\BIBentryALTinterwordstretchfactor}{4}
\providecommand{\BIBentryALTinterwordspacing}{\spaceskip=\fontdimen2\font plus
\BIBentryALTinterwordstretchfactor\fontdimen3\font minus \fontdimen4\font\relax}
\providecommand{\BIBforeignlanguage}[2]{{%
\expandafter\ifx\csname l@#1\endcsname\relax
\typeout{** WARNING: IEEEtran.bst: No hyphenation pattern has been}%
\typeout{** loaded for the language `#1'. Using the pattern for}%
\typeout{** the default language instead.}%
\else
\language=\csname l@#1\endcsname
\fi
#2}}
\providecommand{\BIBdecl}{\relax}
\BIBdecl

\bibitem{randolph2020power}
M.~Randolph \emph{et~al.}, ``Power side-channel attack analysis: A review of 20 years of study for the layman,'' \emph{Cryptography}, 2020.

\bibitem{smart2000physical}
N.~P. Smart, ``Physical side-channel attacks on cryptographic systems,'' \emph{Software Focus}, vol.~1, no.~2, pp. 6--13, 2000.

\bibitem{huss2013amasive}
S.~A. Huss, M.~St{\"o}ttinger, and M.~Zohner, ``Amasive: an adaptable and modular autonomous side-channel vulnerability evaluation framework,'' in \emph{Number Theory and Cryptography: Papers in Honor of Johannes Buchmann on the Occasion of His 60th Birthday}.\hskip 1em plus 0.5em minus 0.4em\relax Springer, 2013, pp. 151--165.

\bibitem{wang2012scare}
X.~Wang, S.~Narasimhan, A.~Krishna, and S.~Bhunia, ``Scare: Side-channel analysis based reverse engineering for post-silicon validation,'' in \emph{2012 25th International Conference on VLSI Design}.\hskip 1em plus 0.5em minus 0.4em\relax IEEE, 2012, pp. 304--309.

\bibitem{hwang2006aes}
D.~D. Hwang, K.~Tiri, A.~Hodjat, B.-C. Lai, S.~Yang, P.~Schaumont, and I.~Verbauwhede, ``Aes-based security coprocessor ic in 0.18-$ muhbox m $ cmos with resistance to differential power analysis side-channel attacks,'' \emph{IEEE Journal of Solid-State Circuits}, vol.~41, no.~4, pp. 781--792, 2006.

\bibitem{schmidt2009probing}
J.-M. Schmidt and C.~H. Kim, ``A probing attack on aes,'' in \emph{Information Security Applications: 9th International Workshop, WISA 2008, Jeju Island, Korea, September 23-25, 2008, Revised Selected Papers 9}.\hskip 1em plus 0.5em minus 0.4em\relax Springer, 2009, pp. 256--265.

\bibitem{becker2013test}
G.~Becker, J.~Cooper, E.~DeMulder, G.~Goodwill, J.~Jaffe, G.~Kenworthy, T.~Kouzminov, A.~Leiserson, M.~Marson, P.~Rohatgi \emph{et~al.}, ``Test vector leakage assessment (tvla) methodology in practice,'' in \emph{International Cryptographic Module Conference}, vol. 1001.\hskip 1em plus 0.5em minus 0.4em\relax sn, 2013, p.~13.

\bibitem{gattu2020power}
N.~Gattu \emph{et~al.}, ``Power side channel attack analysis and detection,'' in \emph{Proceedings of the 39th International Conference on Computer-Aided Design}, 2020.

\bibitem{akkar2001implementation}
M.-L. Akkar and C.~Giraud, ``An implementation of des and aes, secure against some attacks,'' in \emph{Cryptographic Hardware and Embedded Systems—CHES 2001: Third International Workshop Paris, France, May 14--16, 2001 Proceedings 3}.\hskip 1em plus 0.5em minus 0.4em\relax Springer, 2001, pp. 309--318.

\bibitem{blomer2004provably}
J.~Bl{\"o}mer \emph{et~al.}, ``Provably secure masking of aes,'' in \emph{International workshop on selected areas in cryptography}.\hskip 1em plus 0.5em minus 0.4em\relax Springer, 2004.

\bibitem{golic2003multiplicative}
J.~D. Goli{\'c} and C.~Tymen, ``Multiplicative masking and power analysis of aes,'' in \emph{Cryptographic Hardware and Embedded Systems-CHES 2002: 4th International Workshop Redwood Shores, CA, USA, August 13--15, 2002 Revised Papers 4}.\hskip 1em plus 0.5em minus 0.4em\relax Springer, 2003, pp. 198--212.

\bibitem{oswald2005side}
E.~Oswald, S.~Mangard, N.~Pramstaller, and V.~Rijmen, ``A side-channel analysis resistant description of the aes s-box,'' in \emph{Fast Software Encryption: 12th International Workshop, FSE 2005, Paris, France, February 21-23, 2005, Revised Selected Papers 12}.\hskip 1em plus 0.5em minus 0.4em\relax Springer, 2005, pp. 413--423.

\bibitem{messerges2000securing}
T.~S. Messerges, ``Securing the aes finalists against power analysis attacks,'' in \emph{International Workshop on Fast Software Encryption}.\hskip 1em plus 0.5em minus 0.4em\relax Springer, 2000, pp. 150--164.

\bibitem{belaid2015masking}
S.~Bela{\"\i}d, V.~Grosso, and F.-X. Standaert, ``Masking and leakage-resilient primitives: One, the other (s) or both?'' \emph{Cryptography and Communications}, vol.~7, pp. 163--184, 2015.

\bibitem{balasch2015cost}
J.~Balasch, B.~Gierlichs, V.~Grosso, O.~Reparaz, and F.-X. Standaert, ``On the cost of lazy engineering for masked software implementations,'' in \emph{Smart Card Research and Advanced Applications: 13th International Conference, CARDIS 2014, Paris, France, November 5-7, 2014. Revised Selected Papers 13}.\hskip 1em plus 0.5em minus 0.4em\relax Springer, 2015, pp. 64--81.

\bibitem{tempelmeier2016maskver}
M.~Tempelmeier and G.~Sigl, ``Maskver: a tool helping designers detect flawed masking implementations,'' in \emph{2016 1st IEEE International Verification and Security Workshop (IVSW)}.\hskip 1em plus 0.5em minus 0.4em\relax IEEE, 2016, pp. 1--6.

\bibitem{moos2019glitch}
T.~Moos, A.~Moradi, T.~Schneider, and F.-X. Standaert, ``Glitch-resistant masking revisited: Or why proofs in the robust probing model are needed,'' \emph{IACR Transactions on Cryptographic Hardware and Embedded Systems}, pp. 256--292, 2019.

\bibitem{knichel2021automated}
D.~Knichel, A.~Moradi, N.~M{\"u}ller, and P.~Sasdrich, ``Automated generation of masked hardware,'' \emph{Cryptology ePrint Archive}, 2021.

\bibitem{mangard2003simple}
S.~Mangard, ``A simple power-analysis (spa) attack on implementations of the aes key expansion,'' in \emph{Information Security and Cryptology—ICISC 2002: 5th International Conference Seoul, Korea, November 28--29, 2002 Revised Papers 5}.\hskip 1em plus 0.5em minus 0.4em\relax Springer, 2003, pp. 343--358.

\bibitem{kocher1999differential}
P.~Kocher, J.~Jaffe, and B.~Jun, ``Differential power analysis,'' in \emph{Advances in Cryptology—CRYPTO’99: 19th Annual International Cryptology Conference Santa Barbara, California, USA, August 15--19, 1999 Proceedings 19}.\hskip 1em plus 0.5em minus 0.4em\relax Springer, 1999, pp. 388--397.

\bibitem{kocher2011introduction}
P.~Kocher, J.~Jaffe, B.~Jun, and P.~Rohatgi, ``Introduction to differential power analysis,'' \emph{Journal of Cryptographic Engineering}, vol.~1, pp. 5--27, 2011.

\bibitem{brier2004correlation}
E.~Brier, C.~Clavier, and F.~Olivier, ``Correlation power analysis with a leakage model,'' in \emph{Cryptographic Hardware and Embedded Systems-CHES 2004: 6th International Workshop Cambridge, MA, USA, August 11-13, 2004. Proceedings 6}.\hskip 1em plus 0.5em minus 0.4em\relax Springer, 2004, pp. 16--29.

\bibitem{fang2015balance}
X.~Fang, P.~Luo, Y.~Fei, and M.~Leeser, ``Balance power leakage to fight against side-channel analysis at gate level in fpgas,'' in \emph{2015 IEEE 26th International Conference on Application-specific Systems, Architectures and Processors (ASAP)}.\hskip 1em plus 0.5em minus 0.4em\relax IEEE, 2015, pp. 154--155.

\bibitem{veyrat2012shuffling}
N.~Veyrat-Charvillon, M.~Medwed, S.~Kerckhof, and F.-X. Standaert, ``Shuffling against side-channel attacks: A comprehensive study with cautionary note,'' in \emph{Advances in Cryptology--ASIACRYPT 2012: 18th International Conference on the Theory and Application of Cryptology and Information Security, Beijing, China, December 2-6, 2012. Proceedings 18}.\hskip 1em plus 0.5em minus 0.4em\relax Springer, 2012, pp. 740--757.

\bibitem{damgaard2010perfectly}
I.~Damg{\aa}rd, Y.~Ishai, and M.~Kr{\o}igaard, ``Perfectly secure multiparty computation and the computational overhead of cryptography,'' in \emph{Annual international conference on the theory and applications of cryptographic techniques}.\hskip 1em plus 0.5em minus 0.4em\relax Springer, 2010, pp. 445--465.

\bibitem{morrison2014synthesis}
M.~Morrison and N.~Ranganathan, ``Synthesis of dual-rail adiabatic logic for low power security applications,'' \emph{IEEE Transactions on Computer-Aided Design of Integrated Circuits and Systems}, vol.~33, no.~7, pp. 975--988, 2014.

\bibitem{he2019rtl}
M.~He \emph{et~al.}, ``Rtl-psc: Automated power side-channel leakage assessment at register-transfer level,'' in \emph{IEEE VTS}, 2019.

\bibitem{10508974}
A.~Srivastava, S.~Das, N.~Choudhury, R.~Psiakis, P.~H. Silva, D.~Pal, and K.~Basu, ``Scar: Power side-channel analysis at rtl level,'' \emph{IEEE Transactions on Very Large Scale Integration (VLSI) Systems}, pp. 1--14, 2024.

\bibitem{wolf2016yosys}
C.~Wolf, ``Yosys open synthesis suite,'' 2016.

\bibitem{bertsimas1993simulated}
D.~Bertsimas and J.~Tsitsiklis, ``Simulated annealing,'' \emph{Statistical science}, vol.~8, no.~1, pp. 10--15, 1993.

\bibitem{kirkpatrick1983optimization}
S.~Kirkpatrick, C.~D. Gelatt~Jr, and M.~P. Vecchi, ``Optimization by simulated annealing,'' \emph{science}, vol. 220, no. 4598, pp. 671--680, 1983.

\bibitem{karp1990optimal}
R.~M. Karp, U.~V. Vazirani, and V.~V. Vazirani, ``An optimal algorithm for on-line bipartite matching,'' in \emph{Proceedings of the twenty-second annual ACM symposium on Theory of computing}, 1990, pp. 352--358.

\bibitem{fukuda1992finding}
K.~Fukuda and T.~Matsui, ``Finding all minimum-cost perfect matchings in bipartite graphs,'' \emph{Networks}, vol.~22, no.~5, pp. 461--468, 1992.

\bibitem{formality2010equivalence}
S.~Formality, ``Equivalence checking using,'' 2010.

\bibitem{UMC65nm}
{United Microelectronics Corporation (UMC)}, ``55/65/90nm technologies,'' \url{https://www.umc.com/en/Product/technologies/Detail/55_65_90nm}, accessed: 2024-04-29.

\bibitem{Nangate45}
T.~O. Project, ``Openroad-flow-scripts: Nangate45 platform,'' \url{https://github.com/The-OpenROAD-Project/OpenROAD-flow-scripts/tree/master/flow/platforms/nangate45}, 2024, accessed: 2024-04-29.

\bibitem{martins2015open}
M.~Martins, J.~M. Matos, R.~P. Ribas, A.~Reis, G.~Schlinker, L.~Rech, and J.~Michelsen, ``Open cell library in 15nm freepdk technology,'' in \emph{Proceedings of the 2015 Symposium on International Symposium on Physical Design}, 2015, pp. 171--178.

\bibitem{aes}
``Verilog designs,'' \url{http://www.aoki.ecei.tohoku.ac.jp/crypto/web/cores.html}, 2024.

\bibitem{present}
T.~De~Cnudde \emph{et~al.}, ``Higher-order glitch resistant implementation of the present s-box,'' in \emph{BalkanCryptSec}.\hskip 1em plus 0.5em minus 0.4em\relax Springer, 2015.

\bibitem{imran2021design}
M.~Imran \emph{et~al.}, ``Design space exploration of saber in 65nm asic,'' in \emph{Proceedings of the 5th Workshop on Attacks and Solutions in Hardware Security}, 2021.

\bibitem{yaman2021hardware}
F.~Yaman \emph{et~al.}, ``A hardware accelerator for polynomial multiplication operation of crystals-kyber pqc scheme,'' in \emph{2021 IEEE DATE}.

\bibitem{ding2018towards}
A.~A. Ding, L.~Zhang, F.~Durvaux, F.-X. Standaert, and Y.~Fei, ``Towards sound and optimal leakage detection procedure,'' in \emph{Smart Card Research and Advanced Applications: 16th International Conference, CARDIS 2017, Lugano, Switzerland, November 13--15, 2017, Revised Selected Papers}.\hskip 1em plus 0.5em minus 0.4em\relax Springer, 2018, pp. 105--122.

\bibitem{ding2016simpler}
A.~A. Ding, C.~Chen, and T.~Eisenbarth, ``Simpler, faster, and more robust t-test based leakage detection,'' in \emph{Constructive Side-Channel Analysis and Secure Design: 7th International Workshop, COSADE 2016, Graz, Austria, April 14-15, 2016, Revised Selected Papers 7}.\hskip 1em plus 0.5em minus 0.4em\relax Springer, 2016, pp. 163--183.

\bibitem{schneider2016leakage}
T.~Schneider and A.~Moradi, ``Leakage assessment methodology: Extended version,'' \emph{Journal of Cryptographic Engineering}, vol.~6, pp. 85--99, 2016.

\bibitem{ferrufino2023fobos}
E.~Ferrufino, L.~Beckwith, A.~Abdulgadir, and J.-P. Kaps, ``Fobos 3: An open-source platform for side-channel analysis and benchmarking,'' in \emph{Proceedings of the 2023 Workshop on Attacks and Solutions in Hardware Security}, 2023, pp. 5--14.

\bibitem{masking}
``A very compact ``perfectly masked” s-box for aes,'' \url{https://faculty.nps.edu/drcanrig/pub/acns2008corr.pdf}.

\bibitem{shahmirzadi2021new}
A.~R. Shahmirzadi, D.~Bo{\v{z}}ilov, and A.~Moradi, ``New first-order secure aes performance records,'' \emph{Cryptology ePrint Archive}, 2021.

\bibitem{askeland2022guarding}
A.~Askeland, S.~Dhooghe, S.~Nikova, V.~Rijmen, and Z.~Zhang, ``Guarding the first order: The rise of aes maskings,'' in \emph{International Conference on Smart Card Research and Advanced Applications}.\hskip 1em plus 0.5em minus 0.4em\relax Springer, 2022, pp. 103--122.

\bibitem{de2016masking}
T.~De~Cnudde, O.~Reparaz, B.~Bilgin, S.~Nikova, V.~Nikov, and V.~Rijmen, ``Masking aes with shares in hardware,'' in \emph{International Conference on Cryptographic Hardware and Embedded Systems}.\hskip 1em plus 0.5em minus 0.4em\relax Springer, 2016, pp. 194--212.

\bibitem{bilgin2015trade}
B.~Bilgin, B.~Gierlichs, S.~Nikova, V.~Nikov, and V.~Rijmen, ``Trade-offs for threshold implementations illustrated on aes,'' \emph{IEEE Transactions on Computer-Aided Design of Integrated Circuits and Systems}, vol.~34, no.~7, pp. 1188--1200, 2015.

\bibitem{wegener2018first}
F.~Wegener and A.~Moradi, ``A first-order sca resistant aes without fresh randomness,'' in \emph{Constructive Side-Channel Analysis and Secure Design: 9th International Workshop, COSADE 2018, Singapore, April 23--24, 2018, Proceedings 9}.\hskip 1em plus 0.5em minus 0.4em\relax Springer, 2018, pp. 245--262.

\bibitem{shahmirzadi2021re}
A.~R. Shahmirzadi and A.~Moradi, ``Re-consolidating first-order masking schemes: Nullifying fresh randomness,'' \emph{IACR Transactions on Cryptographic Hardware and Embedded Systems}, pp. 305--342, 2021.

\bibitem{sugawara20193}
T.~Sugawara, ``3-share threshold implementation of aes s-box without fresh randomness,'' \emph{IACR Transactions on Cryptographic Hardware and Embedded Systems}, pp. 123--145, 2019.

\bibitem{yahya2009aes}
A.~Yahya, A.~M. Abdalla, H.~Arabnia, and K.~Daimi, ``An aes-based encryption algorithm with shuffling.'' in \emph{Security and Management}, 2009, pp. 113--116.

\bibitem{wang2013area}
Y.~Wang and Y.~Ha, ``An area-efficient shuffling scheme for aes implementation on fpga,'' in \emph{2013 IEEE International Symposium on Circuits and Systems (ISCAS)}.\hskip 1em plus 0.5em minus 0.4em\relax IEEE, 2013, pp. 2577--2580.

\bibitem{ascad}
``Ascad: Anssi sca database,'' \url{https://github.com/ANSSI-FR/ASCAD}.

\end{thebibliography}

\end{document}